\documentclass[12pt]{article}
\usepackage{amssymb}
\usepackage{epsfig}

\usepackage{graphicx}

\def\L{\mathcal L}
\def\e{\varepsilon}

\newcommand{\wt}{\widetilde}

\begin{document}

\def\a{\alpha}
\def\b{\beta}
\def\c{\chi}
\def\d{\delta}
\def\e{\epsilon}
\def\f{\phi}
\def\g{\gamma}
\def\h{\eta}
\def\i{\iota}
\def\j{\psi}
\def\k{\kappa}
\def\l{\lambda}
\def\m{\mu}
\def\n{\nu}
\def\o{\omega}
\def\p{\pi}
\def\q{\theta}
\def\r{\rho}
\def\s{\sigma}
\def\t{\tau}
\def\u{\upsilon}
\def\x{\xi}
\def\z{\zeta}
\def\D{\Delta}
\def\F{\Phi}
\def\G{\Gamma}
\def\J{\Psi}
\def\L{\Lambda}
\def\O{\Omega}
\def\P{\Pi}
\def\Q{\Theta}
\def\S{\Sigma}
\def\U{\Upsilon}
\def\X{\Xi}

\def\ve{\varepsilon}
\def\vf{\varphi}
\def\vr{\varrho}
\def\vs{\varsigma}
\def\vq{\vartheta}

\def\dg{\dagger}                                     
\def\ddg{\ddagger}                                   
\def\wt#1{\widetilde{#1}}                    
\def\mt{\widetilde{m}_1}
\def\mti{\widetilde{m}_i}
\def\rt{\widetilde{r}_1}
\def\mtt{\widetilde{m}_2}
\def\mttt{\widetilde{m}_3}
\def\rtt{\widetilde{r}_2}
\def\mb{\overline{m}}
\def\VEV#1{\left\langle #1\right\rangle}        
\def\be{\begin{equation}}
\def\ee{\end{equation}}
\def\ds{\displaystyle}
\def\ra{\rightarrow}

\def\bea{\begin{eqnarray}}
\def\eea{\end{eqnarray}}
\def\NO{\nonumber}
\def\Bar#1{\overline{#1}}


\def\pl#1#2#3{Phys.~Lett.~{\bf B {#1}} ({#2}) #3}
\def\np#1#2#3{Nucl.~Phys.~{\bf B {#1}} ({#2}) #3}
\def\prl#1#2#3{Phys.~Rev.~Lett.~{\bf #1} ({#2}) #3}
\def\pr#1#2#3{Phys.~Rev.~{\bf D {#1}} ({#2}) #3}
\def\zp#1#2#3{Z.~Phys.~{\bf C {#1}} ({#2}) #3}
\def\cqg#1#2#3{Class.~and Quantum Grav.~{\bf {#1}} ({#2}) #3}
\def\cmp#1#2#3{Commun.~Math.~Phys.~{\bf {#1}} ({#2}) #3}
\def\jmp#1#2#3{J.~Math.~Phys.~{\bf {#1}} ({#2}) #3}
\def\ap#1#2#3{Ann.~of Phys.~{\bf {#1}} ({#2}) #3}
\def\prep#1#2#3{Phys.~Rep.~{\bf {#1}C} ({#2}) #3}
\def\ptp#1#2#3{Progr.~Theor.~Phys.~{\bf {#1}} ({#2}) #3}
\def\ijmp#1#2#3{Int.~J.~Mod.~Phys.~{\bf A {#1}} ({#2}) #3}
\def\mpl#1#2#3{Mod.~Phys.~Lett.~{\bf A {#1}} ({#2}) #3}
\def\nc#1#2#3{Nuovo Cim.~{\bf {#1}} ({#2}) #3}
\def\ibid#1#2#3{{\it ibid.}~{\bf {#1}} ({#2}) #3}

\title{
\vspace*{15mm}
\bf Viability of Dirac phase leptogenesis}
\author{{\Large Alexey Anisimov$^a$, Steve Blanchet$^b$ and
Pasquale Di Bari$^b$}
\\
$^a${\it Institut de Th\'eorie des Ph\'{e}nom\`{e}nes Physiques} \\
{\it Ecole Polytechnique F\'{e}d\'{e}rale de Lausanne} \\
{\it CH-1015 Lausanne, Switzerland} \\
$^b${\it Max-Planck-Institut f\"{u}r Physik} \\
{\it (Werner-Heisenberg-Institut)} \\
{\it F\"{o}hringer Ring 6, 80805 M\"{u}nchen, Germany}}

\maketitle \thispagestyle{empty}


\begin{abstract}
We discuss the conditions for a non-vanishing Dirac phase $\d$ and
mixing angle $\theta_{13}$, sources of $C\!P$ violation in
neutrino oscillations, to be uniquely responsible for the observed
matter-antimatter asymmetry of the universe through
leptogenesis. We show that this scenario, that we call
$\d$-leptogenesis, is viable when the degenerate limit (DL)
for the heavy right-handed (RH) neutrino spectrum is considered. We derive
an interesting joint condition on $\sin\theta_{13}$ and the absolute neutrino
mass scale that can be tested in future neutrino oscillation experiments.
In the limit of hierarchical heavy RH neutrino spectrum (HL), we strengthen
the previous result that $\d$-leptogenesis is only very marginally allowed,
even when the production from the two heavier RH neutrinos is taken into account.
An improved experimental upper bound on $\sin\theta_{13}$ and (or) an account of
quantum kinetic effects could completely rule out this option in the future.
Therefore, $\d$-leptogenesis can be also regarded as a motivation for
models with degenerate heavy neutrino spectrum.
\end{abstract}

\section{Introduction}

Leptogenesis \cite{fy}, a cosmological consequence of the
see-saw mechanism \cite{seesaw}, provides an attractive
explanation for  the baryon asymmetry of the Universe,
one of the most long-standing cosmological puzzles.
A lepton asymmetry is produced in the decays of the very heavy
RH neutrinos predicted by the see-saw mechanism.
In order for the ($B-L$ conserving) sphaleron processes to be able
to convert part of the lepton asymmetry into a baryon asymmetry,
very high temperatures,
$T\gtrsim M_{\rm ew}\sim 100\,{\rm GeV}$,
are required in the early Universe \cite{sphalerons}.

In comparison with other models of baryogenesis, leptogenesis
offers the unique advantage of relying on an ingredient of
physics beyond the Standard Model, neutrino masses, already confirmed
by the experiments. Furthermore and very interestingly,
a quantitative analysis \cite{window} shows that the values of the
atmospheric and of the solar neutrino mass scales,
inferred from neutrino mixing experiments,
favor leptogenesis to work in a mildly `strong wash-out regime':
inverse processes are strong enough to wash-out any contribution
to the final asymmetry depending on the initial conditions but not
too strong to prevent successful leptogenesis.
In this way the observed matter-antimatter asymmetry can be
unambiguously explained within a minimal
extension of the SM where RH neutrinos with a Majorana mass term
and Yukawa couplings are added to the SM Lagrangian and the see-saw
limit is assumed. No particular assumptions on the initial conditions
are required, in complete analogy with what happens in the calculation
of the primordial nuclear abundances within standard Big Bang Nucleosynthesis.

In a typical $N_1$-dominated scenario, the asymmetry is dominantly produced
from the decays of the lightest RH neutrino $N_1$. A
necessary (but not sufficient) condition is the assumption of a
mild hierarchy in the heavy neutrino mass spectrum, such that $M_2$,
the mass of the next-to-lightest RH neutrino, is
approximately three times larger than $M_1$, the mass
of the lightest RH neutrino \cite{beyond}.

In an unflavored analysis, a stringent lower bound on $M_1$ holds
\cite{di}. In the decoupling limit, when the $N_1$-decay parameter
$K_1\rightarrow 0$ and assuming an initial thermal abundance, one
finds \footnote{More exactly, in \cite{cmb}, it was found $4\times
10^8\,{\rm GeV}$. Here we are using a slightly higher value that
is obtained, as we will see, when the reduced experimental error
on the baryon asymmetry and on the atmospheric neutrino mass scale
is taken into account.} $M_1\gtrsim 5\times 10^8\,{\rm GeV}$
\cite{cmb}. However, there are different drawbacks for the
saturation of this lower bound that is anyway strongly model
dependent. A more significant and stringent lower bound,
$M_1\gtrsim 3\times 10^9\,{\rm GeV}$, is obtained at the onset of
the strong wash-out regime, $K_1\simeq K_{\star}\simeq 3.3$
\cite{annals,flavorlep}, where the final asymmetry does not depend
on the initial conditions. This lower bound implies an associated
lower bound \cite{annals,giudice}, $T_{\rm reh}\gtrsim 1.5\times
10^{9}\,{\rm GeV}$, on the value of the temperature at the
beginning of the standard radiation dominated regime, the
reheating temperature within inflation.

As we said, the assumption of a mild hierarchy in the heavy
neutrino mass spectrum is not sufficient to guarantee that
the $N_1$-dominated scenario holds.
It is indeed possible that, for a proper choice of the see-saw
parameters, a $N_2$-dominated scenario holds, where the final asymmetry
is dominated by the contribution from the decays of the next-to-lightest
RH neutrino $N_2$ \cite{geometry}.
In this case the lower bound on $M_1$ does not hold
any more and is replaced by a lower bound on $M_2$, however still
implying a lower bound on $T_{\rm reh}$.

Even when flavor effects \cite{bcst,endo,pilaund,nardi,abada,abada2}
are taken into account
\footnote{
Flavor effects were first considered in \cite{bcst},
and then in \cite{endo} in the particular case of 2
RH neutrinos. However, in these papers, it was found
that flavor effects can only induce small corrections
to the final asymmetry compared to the unflavored case.
The possibility for a large enhancement was first found
in \cite{pilaund} in the case of resonant leptogenesis
and more generally in \cite{nardi,abada,abada2}, where the
typical factor 2--3 enhancement of the final asymmetry 
induced by flavor effects was also first understood.
As for the potential role of low energy phases 
in providing an additional source of 
$C\!P$ violation relevant for leptogenesis, it was first 
discussed in \cite{nardi}.},
these lower bounds do not get relaxed \cite{flavorlep,abada3}.
In particular, flavor effects do not help to alleviate
the conflict with the upper bound on the reheating temperature coming from the
avoidance of the gravitino problem when a supersymmetric framework
is considered \cite{gravitino}. On the other hand, flavor effects relax the
lower bound on $M_1$ for $K_1\gg K_{\star}$ \cite{abada} and, interestingly, it has been shown
that the Dirac phase and, more efficiently, the Majorana phases
can strongly enhance the amount of the relaxation \cite{flavorlep}.

Besides this effect, there is an even more interesting role played by
the Majorana and Dirac phases when flavor effects are taken into account.
In an unflavored analysis, the final asymmetry does not
depend on the low-energy phases and this represents a limit to the possibility
of further tightening the link between leptogenesis predictions and
low-energy neutrino experiments.

In \cite{nardi} it has been shown that, accounting for
flavor effects, an asymmetry can be generated even when the
total $C\!P$ asymmetry vanishes. This is possible because flavor
effects introduce an additional source of $C\!P$ asymmetry stemming
from low-energy phases. The flavor composition
of the anti-lepton produced in the decay of the RH neutrino
can be indeed different from the one of the $C\!P$ conjugated lepton.
In this way a new intriguing scenario arises, where
the Majorana and the Dirac phases, potentially observable in low-energy
neutrino experiments, could act as the unique source of $C\!P$ violation
responsible for the observed matter-antimatter asymmetry of the Universe.
First calculations have been presented in \cite{abada2} for particular
values of $M_1$ and within a two RH neutrino scenario, corresponding
to a specific choice of the see-saw orthogonal matrix \cite{casas}.

In \cite{flavorlep} it has been first shown that successful
leptogenesis stemming only from low-energy phases is possible and
the lower bound on $M_1$, with its dependence on $K_1$ and on the
initial conditions, has been calculated for a specific choice of the
see-saw orthogonal matrix in the HL.
It has been found that for values of $K_1$ in the strong wash-out regime,
the allowed region is very constrained when just Majorana phases are
switched on, and it is even worse when only the Dirac phase is switched on,
even for $\sin\theta_{13}$ close to its experimental upper bound.
Compared to the usual cases where high-energy phases
contribute to $C\!P$ violation as well, the lower bounds on $M_1$
and on the reheating temperature get much more stringent,
especially in the strong wash-out regime and in particular for
values of $K_1$ in the range favored by neutrino mixing experiments.
Therefore, one can say that the asymmetry production
from low-energy phases is somehow secondary compared to the
usual case when leptogenesis proceeds from the high-energy phases
contained in the see-saw orthogonal matrix.
This conclusion has been recently confirmed also in \cite{antusch}
in the context of MSSM. In \cite{pascoli,pascoli2,branco}, the
results have been generalized for an arbitrary choice
of the orthogonal matrix and of the low-energy phases
but without a study of
the dependence on $K_1$ and on the initial conditions.

In this paper we focus on the particularly interesting case of
$\d$-leptogenesis, where the Dirac phase, which has realistic chances
to be observed in neutrino mixing experiments for not too small values of
$\sin\theta_{13}$, is the only non-vanishing phase.
We study the dependence of the final asymmetry on the initial conditions,
finding the onset of the strong wash-out regime and showing the dependence
of the $M_1$ lower bound on the important decay parameter $K_1$, whose
value is related to the
values of the neutrino masses and at the same time determines the
efficiency of the asymmetry production (involving both the
production of the heavy neutrinos and the wash-out).
We first obtain that, in the HL, the possibility to explain the observed
asymmetry is only marginally allowed and just limited to the
less relevant weak wash-out regime, when the correct condition
for the validity of the fully flavored regime is taken into account \cite{zeno}.
Then we point out that this obstacle can be nicely
circumvented going beyond the HL.
Indeed, like in the unflavored case
\cite{crv,pilaftsis,beyond}, the flavored $C\!P$ asymmetries, and
consequently the final $B-L$ asymmetry, get enhanced and the
lower bounds on $M_1$ and on $T_{\rm reh}$ get relaxed.
The possibility of $\d$-leptogenesis beyond the HL has been already
studied in \cite{pascoli2} within resonant leptogenesis \cite{pilaftsis},
where the heavy neutrino mass differences are equal to the resonance widths,
for initial vanishing abundance and in \cite{selma} in the
context of radiative leptogenesis \cite{radiative}
with minimal flavor violation \cite{MFV}.

 We perform a general analysis in the degenerate limit (DL),
where at least one of the degeneracies $\d_{ji}\equiv (M_j-M_i)/M_i\lesssim 0.01$.
We show that in this case the strong wash-out regime always holds
and the lower bound on $M_1$ can be expressed through $\d_{ji}$ and the quantity
$\D\equiv\sin\theta_{13}\,\sin\d$. In the most extreme case
of resonant leptogenesis this turns both into a lower
bound on $\theta_{13}$ and into an upper bound on the absolute
neutrino mass scale that depend on each other.
In this way we find that $\d$-leptogenesis can indeed explain the observed
matter-antimatter asymmetry of the Universe in the strong
wash-out regime and therefore, like leptogenesis from high-energy phases,
exhibits the same virtue of independence of the initial conditions.

In Section 2 we introduce the general framework and set the notation.
In Section 3 we present the results in the HL. We confirm, in a more
general way, the conclusions of \cite{flavorlep}, showing that the allowed
region for $\d$-leptogenesis is quite restricted, especially in the
strong wash-out regime and considering that the asymmetry production
has to switch off for $M_1\gtrsim 10^{12}\,{\rm GeV}$,
when the unflavored case is recovered and $C\!P$ violation from low-energy
phases turns off. We also verify that this conclusion holds even when the
asymmetry production from the two heavier RH neutrinos is taken into account.
On the other hand, we show that
a $N_2$-dominated scenario can also be realized in $\d$-leptogenesis.

We conclude that one needs to go beyond the HL for successful $\d$-leptogenesis
in the strong wash-out regime and in any case not to be just marginally allowed.
Therefore, in Section 4 we study the DL
showing that successful $\d$-leptogenesis is possible and we
find a condition that relates $\d_{ji}$ to $M_i$ ($j=2,3$ and $i=1,2$)
and to $\D$. We also find an upper bound on the absolute neutrino mass
scale dependent on $\sin\theta_{13}$ that makes $\d$-leptogenesis
falsifiable independently of the RH neutrino spectrum.
In Section 5 we draw the conclusions.

\section{General framework}

Adding to the Standard Model three RH neutrinos
with a Majorana mass term $M$ and Yukawa couplings $h$, after
spontaneous breaking a Dirac mass term, $m_D=v\,h$,
is generated by the vev $v$ of the Higgs boson.
In the see-saw limit, $M\gg m_D$, the spectrum of neutrino mass eigenstates
splits in two sets, a very heavy one, $N_1,N_2$ and $N_3$ with masses
respectively $M_1\leq M_2 \leq M_3$ and
almost coinciding with
the eigenvalues of $M$, and a light one, with masses $m_1\leq m_2\leq m_3$
corresponding to the eigenvalues of the light neutrino
mass matrix given by the see-saw formula \cite{seesaw},
\be
m_{\nu}= - m_D {1\over M} m_D^T \, .
\ee
Neutrino mixing experiments measure two light neutrino mass
squared differences. In a normal scheme one has
$m^{\,2}_3-m_2^{\,2}=\Delta m^2_{\rm atm}$ and
$m^{\,2}_2-m_1^{\,2}=\Delta m^2_{\rm sol}$,
whereas in an inverted scheme one has
$m^{\,2}_3-m_2^{\,2}=\Delta m^2_{\rm sol}$
and $m^{\,2}_2-m_1^{\,2}=\Delta m^2_{\rm atm}$.
For $m_1\gg m_{\rm atm} \equiv
\sqrt{\Delta m^2_{\rm atm}+\Delta m^2_{\rm sol}}=
(0.052\pm 0.002)\,{\rm eV}$ \cite{concha}
the spectrum is quasi-degenerate, while for
$m_1\ll m_{\rm sol}\equiv \sqrt{\D m^2_{\rm sol}}
=(0.0089\pm 0.0002)\,{\rm eV}$ \cite{concha} is fully hierarchical.

In the early Universe, the decays of the heavy neutrinos into
leptons and Higgs bosons produce, in general, a lepton number
that is partly converted into a baryon number by sphaleron
($B-L$ conserving) processes if the temperature is higher than about
$100\,{\rm GeV}$ \cite{sphalerons}.

An important role is played by the decay parameters of the heavy
neutrinos  defined as $K_i\equiv \widetilde{\G}_i/H_{T=M_i}$,
the ratios of the decay widths to the expansion rate
when the RH neutrinos start to become non-relativistic
at $T=M_i$. For $K_i \ll 1$ the bulk of the $N_i$ decays occurs when they
are non-relativistic and the inverse decays are not effective anymore.
In this case all decays occur out-of-equilibrium and
the wash-out of the asymmetry is weak. On the other hand, for
$K_i\gg 1$, the heavy neutrinos decays are balanced by inverse processes.
In this case the heavy neutrino abundance tracks quite closely the equilibrium
abundance and the wash-out of the asymmetry is potentially, but not necessarily,
strong. The answer depends on a detailed description of flavor effects
that are triggered by the charged lepton Yukawa interactions with a
rate $\G_{\alpha}\simeq 5\times 10^{-3}\,T\,f^2_{\alpha}\,(\a=e,\m,\t)$
~\cite{Campbell:1992jd}, where the $f_{\alpha}$'s are the charged lepton
Yukawa couplings in the diagonal basis.

If $\G_{\alpha}\ll \sum_{i}\,\G_{\rm ID}^i$
\footnote{Notice that more rigorously
this condition should be written replacing the simple sum of
the inverse decays rates with a sum weighted with
projectors taking into account that the lepton produced by
the decay of a RH neutrino $N_i$ is different by that lepton produced by
the decay of a RH neutrino $N_{j\neq i}$ and therefore is not in general fully
absorbed by the $N_{j\neq i}$ inverse decay \cite{bcst}.},
during all the relevant period of the asymmetry generation, then
the lepton state coherence is preserved between decays and inverse
decays and the unflavored regime, where flavor effects
are negligible, holds.  This requirement implies \cite{zeno}
\be\label{unflavored}
M_1\gtrsim 5\times 10^{11}\,{\rm GeV} \, .
\ee
In the unflavored regime the condition $K_i\gg 1$ is also
sufficient for the wash-out regime to be strong.
It is important to stress that in this regime the only source of
$C\!P$ violation is due to a different total decay rate into
leptons and anti-leptons and, as it is well known, it stems
uniquely from high-energy phases. Therefore, in the unflavored regime,
$\d$-leptogenesis is not viable.

If the charged lepton Yukawa interactions are in equilibrium
($\G_{\a}> H$) and faster than inverse decays,
\be\label{condition}
\G_{\a}\gtrsim \sum_{i}\,\G_{\rm ID}^i \, ,
\ee
during the relevant period of the asymmetry generation,
then the lepton quantum states lose coherence between the production
at decay and the subsequent absorption in inverse processes.
 In this way the Higgs bosons interact incoherently with leptons of each
flavor. In the limit case, when the quantum state becomes completely incoherent
and is fully projected in one of the flavor eigenstates, each lepton flavor
can be treated as a statistically independent particle species
and a `fully flavored regime' is obtained.
One has to distinguish a two-flavor regime, for
$M_1\gtrsim 10^{\rm 9}\,{\rm GeV}$, such that
the condition Eq. (\ref{condition}) is in any case satisfied
only for $\a=\t$, and a three-flavor regime, where the
condition Eq. (\ref{condition}) applies also to $\a=\m$.

In the fully (two or three) flavored regime,  classic Boltzmann equations
can be used like in the unflavored regime, with the difference, in general,
that now each single flavor asymmetry has to be tracked independently.

In the fully flavored regime there are two new effects
compared  to the unflavored regime \cite{nardi}. These can
be understood introducing the projectors  and writing them
as the sum of two terms,
\bea
P_{i\alpha} & \equiv  &
|\langle l_{i}|l_{\alpha}\rangle |^2  =
P_{i\alpha}^0 + {\Delta P_{i\alpha}\over 2} \\
\bar{P}_{i\alpha}& \equiv &
|\langle \bar{l}'_{i}|\bar{l}_{\alpha}\rangle |^2  =
P_{i\alpha}^0 - {\Delta P_{i\alpha}\over 2} \, .
\eea
The first effect is a reduction of the wash-out compared to the
unflavored regime and is described by the tree level contribution
$P_{i\a}^0=(P_{i\a}+\bar{P}_{i\a})/2$ setting
the fraction of the total asymmetry, produced in $N_i$-decays,
that goes into each single flavor $\a$. In the fully flavored regime,
each single inverse decay involves an independent lepton flavor
eigenstate and therefore does not wash out, in general, as much
asymmetry as that one produced in each single decay but an amount
reduced by $P_{i\a}^0$.

The second effect is an additional $C\!P$ violating contribution coming from
a different flavor composition between
$|l_i\rangle$ and $C\!P |\bar{l}_i'\rangle$.
This can be described in terms of the projector differences
$ \D\,P_{i\a}\equiv P_{i\alpha}-\bar{P}_{i\alpha}$,
such that $\sum_{\a}\,\D\,P_{i\a}= 0$.
Indeed, defining the flavored $C\!P$ asymmetries,
\be
\ve_{i\alpha} \equiv
-{\G_{i\alpha}-\overline{\G}_{i\alpha}
\over \G_{i}+\overline{\G}_{i}} \, ,
\ee
where $\G_{i\a}\equiv P^0_{i\a}\,\G_{\a}$ and
$\bar{\G}_{i\a}\equiv P^0_{i\a}\,\bar{\G}_{\a}$,
these can be now written as
\be
\ve_{i\a}=\ve_i\,P^{0}_{i\a}+ {\D\,P_{i\a}\over 2} \, ,
\ee
where $\ve_i\equiv \sum_\a\,{\ve_{i\a}}$ are the total $C\!P$
asymmetries. In the last expression one can see
that the first term is the usual contribution due to
a different decay rate into lepton and anti-leptons and
the second is the additional contribution
due to a possible different flavor composition
between $|l_i\rangle$ and $C\!P |\bar{l}_i'\rangle$.


Taking into account only decays and inverse decays with proper
subtraction of the resonant contribution from $\D L=2$ and $\D L=0$
processes \cite{giudice,pilaund,nardi},
the set of effective classic Boltzmann equations valid in the fully
three-flavored regime can be written as
\bea\label{flke}
{dN_{N_i}\over dz} & = & -D_i\,(N_{N_i}-N_{N_i}^{\rm eq})
\hspace{52mm} (i=1,2,3) \\ \nonumber
{dN_{\D_{\a}}\over dz} & = &
\sum_i\,\ve_{i\a}\,D_i\,(N_{N_i}-N_{N_i}^{\rm eq})
-\sum_i\,P_{i\a}^{0}\, W_i^{\rm ID}\,N_{\D_{\a}} \,
\hspace{5mm} (\a=e,\m,\t) ,
\eea
where $z \equiv M_1/T$ and where we indicated with $N_X$
any particle number or asymmetry $X$ calculated in a portion of co-moving
volume containing one heavy neutrino in ultra-relativistic thermal equilibrium,
so that $N^{\rm eq}_{N_i}(T\gg M_i)=1$.
Defining $x_i\equiv M_i^2/M_1^2$ and $z_i\equiv z\,\sqrt{x_i}$,
the decay factors are given by
\be
D_i \equiv {\G_{D,i}\over H\,z}=K_i\,x_i\,z\,
\left\langle {1\over\gamma_i} \right\rangle   \, ,
\ee
where $H$ is the expansion rate. The total decay rates,
$\G_{D,i}\equiv \G_i+\bar{\G}_i$,
are the product of the decay widths times the
thermally averaged dilation factors
$\langle 1/\gamma_i\rangle$, given by the ratio
${\cal K}_1(z_i)/ {\cal K}_2(z_i)$ of the
modified Bessel functions. The equilibrium abundance and its rate are
also expressed through the modified Bessel functions,
\be
N_{N_i}^{\rm eq}(z_i)= {1\over 2}\,z_i^2\,{\cal K}_2 (z_i) \;\; ,
\hspace{10mm}
{dN_{N_i}^{\rm eq}\over dz_i} =
-{1\over 2}\,z_i^2\,{\cal K}_1 (z_i) \, .
\ee
Finally, the inverse decays wash-out terms are given by
\be\label{WID}
W_i^{\rm ID}(z) =
{1\over 4}\,K_i\,\sqrt{x_i}\,{\cal K}_1(z_i)\,z_i^3 \, .
\ee
We are neglecting the non resonant contributions from
$\D L=2$ and $\D L=0$ processes, a good approximation for
$M_1\ll 10^{14}\,{\rm GeV}\,(m_{\rm atm}^2/\sum_i\,m_i^2)$,
as we will always consider. We are also neglecting
$\D L=1$ scatterings \cite{luty,plum,pilaund2,abada2},
giving a correction to a level less than $\sim 10\% $
\cite{flavorlep} and spectator processes \cite{buchplum,nardi2} that,
at least for a hierarchical heavy neutrino spectrum, produce a correction
to a level less than $\sim 30\% $ \cite{nardi2,abada3}.
In the degenerate limit it cannot be excluded that
the effect of spectator processes is more relevant
and further studies are required. We are also neglecting
thermal corrections \cite{giudice}, that can give relevant (though
with big theoretical uncertainties) corrections in the weak wash-out
regime but negligible ones in the more important strong wash-out regime.

The evolution of the $N_{\D\a}$'s can be worked out in an integral form,
\be
N_{\D\a}(z)=N_{\D\a}^{\rm in}\,
e^{-\sum_i\,P_{i\a}^0\,\int_{z_{\rm in}}^z\,dz'\,W_i^{\rm ID}(z')}
+\sum_i\,\ve_{i\a}\,\k_{i{\a}}(z) \,  ,
\ee
with the 9 efficiency factors given by
\be\label{ef}
\k_{i\a}(z;K_i,P^{0}_{i\a})=-\int_{z_{\rm in}}^z\,dz'\,{dN_{N_i}\over dz'}\,
e^{-\sum_i\,P_{i\a}^0\,\int_{z'}^z\,dz''\,W_i^{\rm ID}(z'')} \,.
\ee
The total final $B-L$ asymmetry is then given by
$N_{B-L}^{\rm f}=\sum_{\a}\,N_{\D_\a}^{\rm f}$. Finally,
assuming a standard thermal history and accounting for the
sphaleron converting coefficient $a_{\rm sph}\sim 1/3$,
the final baryon-to-photon number ratio can be calculated as
\be\label{etaB}
\eta_B=a_{\rm sph}\,{N_{B-L}^{\rm f}\over N_{\gamma}^{\rm rec}}
\simeq 0.96\times 10^{-2}\,N_{B-L}^{\rm f} \, ,
\ee
to be compared with the measured value \cite{WMAP3}
\be\label{etaBobs}
\eta_B^{\rm CMB} = (6.1 \pm 0.2)\times 10^{-10} \, .
\ee
Notice that the efficiency factors depend only on the
$P_{i\a}^0$ but not on the differences $\D P_{i\a}$.
Notice also that, in the two-flavor regime, the individual
electron and muon asymmetries are replaced by one kinetic equation
for the sum, $N_{\D_{e\m}}\equiv N_{\D_{\m}}+N_{\D_e}$,
where the individual flavored $C\!P$ asymmetries and projectors
have also to be replaced by the their sum, namely
$\ve_{1\,e+\m}\equiv \ve_{1\m}+\ve_{1e}$ and $P^0_{1\,e+\m}\equiv
P^0_{1\m}+P^0_{1 e}$ \cite{abada2}.
The calculation is therefore somehow intermediate between the
one-flavor approximation and the three-flavor regime, though the results
are very similar to the three-flavor regime \cite{flavorlep}.

The flavored $C\!P$ asymmetries are given by the following expression \cite{crv}
\be\label{veia}
\ve_{i\a}=
\frac{3}{16 \p (h^{\dag}h)_{ii}} \sum_{j\neq i} \left\{ {\rm Im}\left[h_{\a i}^{\star}
h_{\a j}(h^{\dag}h)_{i j}\right] \frac{\x(x_j/x_i)}{\sqrt{x_j/x_i}}+
\frac{2}{3(x_j/x_i-1)}{\rm Im}
\left[h_{\a i}^{\star}h_{\a j}(h^{\dag}h)_{j i}\right]\right\} \, ,
\ee
where
\be\label{xi}
\xi(x)= {2\over 3}\,x\,
\left[(1+x)\,\ln\left({1+x\over x}\right)-{2-x\over 1-x}\right] \, .
\ee
A parametrization of the Dirac mass matrix,
particularly fruitful within leptogenesis, is obtained
in terms of the see-saw orthogonal matrix $\O$ \cite{casas}
\be\label{Opar}
m_D = U\,D_m^{1/2}\,\O\,D_M^{1/2} \, ,
\ee
where we defined
$D_m\equiv {\rm diag}(m_1,m_2,m_3)$ and
$D_M\equiv {\rm diag}(M_1,M_2,M_3)$.
The matrix $U$ diagonalizes
the light neutrino mass matrix $m_{\nu}$, such that
$U^{\dagger}\,m_{\nu}\,U^{\star}=-D_m$, and it can be identified
with the lepton mixing matrix in a basis where
the charged lepton mass matrix is diagonal. Moreover,
neglecting the effect of the running
of neutrino parameters from high energy to low energy \cite{running},
one can assume that the $U$ matrix can be identified
with the PMNS matrix, partially measured in neutrino
mixing experiments.
For normal hierarchy we adopt the parametrization  \cite{PDG}
\begin{equation}\label{Umatrix}
U=\left( \begin{array}{ccc}
c_{12}\,c_{13} & s_{12}\,c_{13} & s_{13}\,e^{-i\,\d} \\
-s_{12}\,c_{23}-c_{12}\,s_{23}\,s_{13}\,e^{i\,\d} &
c_{12}\,c_{23}-s_{12}\,s_{23}\,s_{13}\,e^{i\,\d} & s_{23}\,c_{13} \\
s_{12}\,s_{23}-c_{12}\,c_{23}\,s_{13}\,e^{i\,\d}
& -c_{12}\,s_{23}-s_{12}\,c_{23}\,s_{13}\,e^{i\,\d}  &
c_{23}\,c_{13}
\end{array}\right)
\times {\rm diag(e^{i\,{\Phi_1\over 2}}, e^{i\,{\Phi_2\over 2}}, 1)}
\, ,
\end{equation}
where $s_{ij}\equiv \sin\theta_{ij}$, $c_{ij}\equiv\cos\theta_{ij}$
and, neglecting the statistical errors, we will use
$\theta_{12}=\pi/6$ and $\theta_{23}=\pi/4$,
compatible with the results from neutrino mixing experiments.
Moreover, we will adopt the $3\s$ range $s_{13}=0-0.20$,
allowed from a global $3\n$ analysis for unitary $U$ \cite{concha},
an approximation that holds
with great precision in the see-saw limit with $M_i\gg 100\,{\rm GeV}$.
Within the convention we are using, $m_1\leq m_2 \leq m_3$,
the case of inverted hierarchy corresponds is obtained performing
a cyclic permutation of columns in the PMNS matrix parametrization
Eq. (\ref{Umatrix}), such that the $i$-th column becomes the $(i+1)$-th.
Since we are interested in understanding whether a non-vanishing Dirac phase
can be the only source of $C\!P$ violation for successful leptogenesis,
we will set the Majorana phases to zero. We will comment later on the
effects of turning on the Majorana phases.

It will also prove useful to introduce the following
parametrization for the see-saw orthogonal matrix,
\be\label{second}
\O({\o}_{21},{\o}_{31},{\o}_{32})
=R_{12}(\o_{21})\,\,
 R_{13}(\o_{31})\,\,
 R_{23}(\o_{32})\,\, ,
\ee
where
\be\label{R}
\mbox{\tiny $
R_{12}=
\left(
\begin{array}{ccc}
 \sqrt{1-{\o}^2_{21}}  &  -{\o}_{21}          & 0 \\
            {\o}_{21}  & \sqrt{1-{\o}^2_{21}} & 0 \\
  0 & 0 & 1
\end{array}
\right) \,
         \,\, , \,\,
R_{13}=
\left(
\begin{array}{ccc}
 \sqrt{1-{\o}^2_{31}}  & 0 &  - {\o}_{31} \\
    0 & 1 & 0 \\
  {\o}_{31} & 0 & \sqrt{1-{\o}^2_{31}}
\end{array}
\right)  \,\, ,
\,\,
R_{23}=
\left(
\begin{array}{ccc}
  1  &  0   & 0   \\
  0  &  \sqrt{1-{\o}^2_{32}} & - {\o}_{32} \\
  0 & {\o}_{32}  & \sqrt{1-{\o}^2_{32}}
\end{array}
\right)$ \, .
}
\ee
Notice that, using the orthogonal parametrization, the
decay parameters $K_i$ can be expressed as linear combinations
of the neutrino masses \cite{fhy,annals}
\be\label{Kimi}
K_i = {\mti\over m_{\star}} = \sum_j\,{m_j\over m_{\star}}\,|\O_{ji}^2| \, ,
\ee
where $\mti\equiv  (m^{\dagger}_D\,m_D)_{ii}/M_i$
are the effective neutrino masses \cite{plum} and where
$m_{\star}$ is the equilibrium neutrino mass \cite{annals} given by
\begin{equation}\label{d}
m_{\star} = {16\, \pi^{5/2}\,\sqrt{g_*} \over 3\,\sqrt{5}}\,
{v^2 \over M_{Pl}} \simeq 1.08\times 10^{-3}\,{\rm eV}\;.
\end{equation}
Barring huge phase cancellations and special forms for $\O$,
typically the $K_i$'s span within the range $[K_{\rm sol},K_{\rm atm}]$
where  $K_{\rm sol}\equiv m_{\rm sol}/m_{\star}=8.2\pm 0.2$
and $K_{\rm atm}\equiv m_{\rm atm}/m_{\star}=48 \pm 2$.

Before entering into a detailed analysis focusing on $\d$-leptogenesis,
we want to discuss some general features concerning the fully flavored
regime and in particular the possibility to have important deviations
from the unflavored case. For definiteness and simplicity, we
refer to the two-flavor case within the $N_1$-dominated scenario,
so that
$N_{B-L}^{\rm f}\simeq \ve_{1\t}\,\k_{1\t}^{\rm f}+
\ve_{1,e+\m}\,\k_{1,e+\m}^{\rm f}$.

Consider first the `democratic case', where $\D P_{1\a}=0$
and $P_{1\t}=P_{1\, e+\m}=1/2$. Summing the two equations
for $\a=\t$ and $\a=e+\m$ one obtains a closed equation for the
total asymmetry where the only effect compared to the unflavored case
is that the wash-out is reduced by a factor two and the final asymmetry
is obtained by replacing $K_1\rightarrow K_1/2$.
Therefore, in the strong wash-out regime ($K_1\gg 1$), since
$\k_{1\a}^{\rm f}\propto K_1^{-1.2}$ \cite{proc},
one has approximately a factor two enhancement.
Let us now consider $P_{1\m}^0< P_{1\t}^0$, still with $\D P_{1\a}=0$.
Since approximately $\k_{1\a}^{\rm f}\propto (P_{1\a}^0)^{-1.2}$ and
at the same time $\ve_{1\a}\propto P_{1\a}^0$, one has that the
final asymmetry stays approximately constant compared to the
democratic case with the two contributions from the $\m$ and $\t$
flavors comparable with each other. Therefore, for vanishing $\D P_{1\a}$,
flavor effects produce just ${\cal O}(1)$ corrections compared to the
unflavored approximation.

This conclusion changes when non-vanishing $\D P_{1\a}$
are considered. In this case there  are two remarkable possibilities.

The {\em first possibility} is the so called one-flavor dominated scenario,
relying on the fact that, for $P^0_{1\a}\rightarrow 0$, one has
${\rm max}(\D P_{1\a})\propto \sqrt{P_{1\a}^0}$ \cite{abada2}.
Therefore, considering now
for example $P_{1\t}\ll P_{1e+\m}\simeq 1$, one has that the asymmetry in
the tauon flavor $\ve_{1\t}\,\k_{1\t}\propto (P_{1\t}^0)^{-0.7}$, showing that
there can be a large enhancement compared to the unflavored case
 in the strong wash-out regime. This brings to a strong relaxation
of the lower bounds on $M_1$ and $T_{\rm reh}$ at $K_1\gg 1$, though,
as we already said, not to a relaxation of the usual lowest bounds at
$K_1\rightarrow 0$ or at $K_1\simeq K_{\star}$. It should also be said
that, as shown in \cite{zeno}, the applicability of the one-flavor dominated
scenario is strongly limited by the condition of validity of the
fully flavored regime Eq. (\ref{condition}).

The {\em second possibility} relies on the observation that,
contrarily to $\ve_1$,
the $\D P_{1\a}$'s depend on the low-energy phases as well and,
even though $\ve_1=0$, they do not vanish if the Dirac or
the Majorana phases do not vanish. Therefore, one can have a final
asymmetry originating just from low-energy phases \cite{nardi}.
This scenario represents, potentially, the most important novelty introduced
by flavor effects compared to the unflavored picture and in the
following Sections we will study it in detail, focusing on the
case of $\d$-leptogenesis, when only the Dirac phase is switched on
while $\O$ is real and the two Majorana phases vanish.

Before concluding this Section, we want to notice that
one can have $\ve_i=0$ not only when
the see-saw orthogonal matrix is real, but also when the absolute neutrino
mass scale increases \cite{abada}. In this way, the low-energy phases can play
an important role in circumventing the upper bound on the neutrino masses
holding in the unflavored regime \cite{bound1,window}.
It is however still to be assessed whether the fully flavored regime
can offer a sufficient description to solve this issue.
In \cite{abada} the bound was found
to be completely nullified by flavor effects. In \cite{zeno} it has been
pointed out how this conclusion relies on a extension of the fully flavored regime
beyond the regime of its validity given by the Eq. (\ref{condition}).
In \cite{riotto} the authors find that in the fully flavored regime, thanks
to spectator processes, the bound holding in the unflavored regime, even though
not nullified, is anyway relaxed  to $m_1\lesssim 2\,{\rm eV}$.

\section{The hierarchical limit}

Let us consider first $\d$-leptogenesis in the HL,
such that $M_3 \gtrsim 3\,M_2 \gtrsim 3\,M_1$ \cite{beyond}.
In the unflavored regime, this assumption typically implies
a $N_1$-dominated scenario, where the final asymmetry is
dominated by the contribution from the lightest RH neutrino decays,
\be\label{N1DS}
N^{\rm f}_{B-L}\simeq \left. N_{B-L}^{\rm f} \right|_{N_1}
\equiv \sum_{\a}\,\ve_{1\a}\,\k_{1\a}\, .
\ee
Indeed, in general, in the HL one has two effects.
The first effect is that the asymmetry production from the two heavier RH
neutrinos, $N_2$ and $N_3$, is typically later on washed out by the
$N_1$ inverse processes and $\k_3^{\rm f},\k_2^{\rm f}\ll \k_1^{\rm f}$.
 The second effect is a consequence of the fact that the total
$C\!P$ asymmetries vanish in the limit when all particles running in the
loops become massless and this yields typically
$|\ve_3|\ll |\ve_2| \ll |\ve_1|$.

However, for a particular choice of the see-saw parameters,
$\O\simeq R_{23}$ and $m_1\lesssim m_{\star}$,
the contribution to the final asymmetry from the next-to-lightest
RH neutrino $N_2$ is not only non-negligible but even dominant, giving rise
to a $N_2$-dominated scenario \cite{geometry}. Indeed
for $\O\simeq R_{23}$ different things happen simultaneously.
First, $N_2$, even though decoupled from $N_1$, is still
coupled to $N_3$ and in the HL the total $C\!P$ asymmetry
$\ve_2$ does not vanish, since it receives a non suppressed contribution
from graphs where $N_3$ runs in the loops.
On the other hand, now one has $\ve_1=0$, since $N_1$
is essentially decoupled from the other two RH neutrinos. At the same time
one also has $K_1\ll 1$, so that the wash-out from $N_1$ inverse processes
is negligible. The final result is that
$|\ve_2\,{\k_2}|\gg |\ve_{i\neq 2}\,\k_{i\neq 2}^{\rm f}|$
and the final asymmetry is dominantly produced from $N_2$-decays.


Therefore, in the unflavored approximation and in the HL,
a condition $w_{32}\simeq 1$ in the
$\O$-matrix parametrization (cf. Eq. (\ref{second}))
is sufficient to have a negligible asymmetry production
from the two heavier RH neutrinos and to guarantee that
the $N_1$-dominated scenario holds.
This condition is even not necessary for
$m_1\gg m_{\star}$, since in this case, due to the
fact that $\mt\geq m_1$, one has necessarily $K_1\gg 1$
and  a wash-out from $N_1$-inverse processes is
anyway strong enough to suppress a possible contribution to the
final asymmetry produced from $N_2$-decays.

When flavor effects are taken into account,
the domain of applicability of the $N_1$-dominated
scenario reduces somehow.
There are two aspects to be considered.

The first aspect is that the wash-out from $N_1$ inverse processes
becomes less efficient. Indeed the projectors $P_{1\a}$ can considerably reduce
the wash-out of the asymmetry produced in the flavor $\a$ from
$N_2$-decays \cite{vives}. This turns the condition
$m_1\gg m_{\star}$ into a looser condition $m_1\gg m_{\star}/P_{1\a}$.
Another effect is that $N_1$ inverse processes can make
part of the asymmetry produced in $N_2$ decays somehow orthogonal
to the the wash-out from $N_1$ inverse processes \cite{bcst,nardi3}.
Recently, it has been also pointed out that spectator processes
can lead to a reduction of the wash-out from $N_1$-inverse processes
as well \cite{shindou}. In this way the assumption
$\k_{2\a}\ll \k_{1\a}$ is not valid in general.

The second aspect concerns the flavored $C\!P$ asymmetries.
In the HL, from the general  expression Eq. (\ref{veia}), one has
\begin{eqnarray} \label{ve1a}
\ve_{1\alpha}&\simeq& \frac{3}{16 \p (h^{\dag}h)_{11}}
\sum_{j\neq 1} \frac{M_1}{M_j} {\rm Im}
\left[h_{\a 1}^{\star} h_{\a j}(h^{\dag}h)_{1 j}\right],\\   \label{ve2a}
\ve_{2\alpha}&\simeq&
\frac{3}{16 \p (h^{\dag}h)_{22}}
\left\{\frac{M_2}{M_3}{\rm Im}
\left[h_{\a 2}^{\star} h_{\a 3}(h^{\dag}h)_{2 3}\right]
-\frac{2}{3} {\rm Im}\left[h_{\a 2}^{\star} h_{\a 1}(h^{\dag}h)_{1 2}\right]
\right\} ,\\
\ve_{3\alpha}&\simeq& -\frac{1}{8\,\p (h^{\dag}h)_{33}}
\sum_{j\neq 3}
\left\{{\rm Im}\left[h_{\a 3}^{\star} h_{\a j}(h^{\dag}h)_{j 3}\right]\right\}.
\end{eqnarray}
Different comments are in order.
The $\ve_{1\a}$'s, like $\ve_1$, vanish for $\O=R_{23}$ while the
$\ve_{2\a}$'s, like $\ve_2$, do not.
On the other hand, in the HL, the $\ve_{2\a}$'s, contrarily to
$\ve_2$, are not suppressed when $\o_{32}=0$ (a particular example is given by
$\O=R_{12}$) but, like $\ve_2$, they vanish for $\O=R_{13}$.

This observation \cite{flavorlep} can also potentially contribute to enlarge
the domain of applicability of the $N_2$-dominated scenario when flavor
effects are taken into account. Another interesting observation is that
the $\ve_{3\a}$'s, contrarily to $\ve_3$, do  not vanish in the HL.
This could open the door even to a $N_3$-dominated
scenario, though  this is possible only for
$M_3\lesssim 10^{12}\,{\rm GeV}$, when flavor effects are
effective in $N_3$ decays.

Therefore, when flavor effects are taken into account,
the conditions of applicability
of the $N_1$-dominated scenario become potentially more restrictive
than in the unflavored case.
There is a clear choice of the parameters,
for $\O=R_{13}$ and $M_3\gtrsim 10^{12}\,{\rm GeV}$,
where the $N_1$-dominated scenario holds. Indeed in this case, in the HL,
one has that $\ve_{2\a}$ and $\ve_3$ are suppressed. This can be considered somehow
opposite to the case $\O=R_{23}$, where the $N_2$-dominated
scenario holds \cite{geometry}.

In general, one can say that the asymmetry produced
from the two heavier RH neutrinos is non-negligible
if two conditions are satisfied.
(i) The asymmetry generated from $N_{2,3}$-decays
at $T\sim M_{2,3}$ has to be non-negligible compared to the
asymmetry generated at $T\sim M_1$ from $N_1$-decays.
This depends on an evaluation of the $C\!P$ asymmetries
$\ve_{2,3}^{\a}$ and of the wash-out due to the same
$N_{2,3}$-inverse processes.
(ii) The asymmetry produced from $N_{2,3}$-decays has not to be
afterwards washed-out by $N_1$-inverse processes.
Notice that this second condition is subordinate to
the first condition.

In the particular case of $\d$-leptogenesis, one has $\ve_2=\ve_3=0$.
This means that the first condition can be satisfied only if
$M_2, M_3\lesssim 10^{12}\,{\rm GeV}$ and
this constitutes already an important limitation.
In the following, we will consider different particular cases, verifying
whether the production from the two heavier RH neutrinos can be neglected
or not. We will find that the situation is actually similar
to what happens in the unflavored case where, except for
the case $\O\sim R_{23}$, a $N_1$-dominated scenario holds.

Let us therefore start showing in detail how to calculate
the contribution to the final asymmetry from $N_1$-decays.
The expression Eq. (\ref{ef}) for the $\k_{1\a}$'s can be
specialized as
\be\label{ef1}
\k_{1\a}(z;K_1,P^{0}_{1\a})=
-\int_{z_{\rm in}}^z\,dz'\,{dN_{N_1}\over dz'}\,
e^{-P_{1\a}^0\,\int_{z'}^z\,dz''\,W_1^{\rm ID}(z'')} \,.
\ee
From the Eq. (\ref{ef}), extending an analytic procedure
derived within the one-flavor approximation \cite{annals},
one can obtain simple analytic expressions for the $\k_{1\a}^{\rm f}$'s.
In the case of an initial thermal abundance ($N_{N_1}^{\rm in}=1$),
defining $K_{1\a}\equiv P^0_{1\a}\,K_1$, one has
\be\label{k1a}
\k_{1\a}^{\rm f} \simeq \k(K_{1\a}) \equiv
{2\over K_{1\a}\,z_B(K_{1\a})}\,
\left(1-e^{-{K_{1\a}\,z_B(K_{1\a})\over 2}}\right) \, ,
\ee
where
\be
z_{B}(K_{1\a}) \simeq 2+4\,K_{1\a}^{0.13}\,e^{-{2.5\over K_{1\a}}} \, .
\ee
In the case of initial vanishing abundance
($N_{N_1}^{\rm in}=0$)
one has to take into account two different contributions,
a negative and a positive one, so that
\be
\k_{1\a}^{\rm f}
=\k_{-}^{\rm f}(K_1,P_{1\a}^{0})+
 \k_{+}^{\rm f}(K_1,P_{1\a}^{0}) \, ,
\ee
whose analytic expressions, used to obtain all presented
results, can be found in \cite{flavorlep}.

The condition for the validity of the fully flavored regime
Eq. (\ref{condition}) can be specialized and re-cast like
\be\label{full}
 M_1\lesssim {10^{12}\,{\rm GeV}\over 2\,W_1^{\rm ID}(z_{B}(K_{1\a}))}.
\ee
This condition neglects the effect of $\D L=1$ scatterings
and of coherent scatterings, the first contributing with inverse decays
to preserve the quantum state coherence, the second, conversely,
in projecting it on the flavor basis \cite{zeno}.
Both of them can be as large as the effect from inverse decays.
Moreover, in a rigorous quantum kinetic description,
it is likely that other subtle effects contribute to the determination of the
exact value of $M_1$ below which the fully flavored regime can be assumed.
Therefore, the condition (\ref{full}) should be regarded as a very qualitative
one. In the plots showing the $M_1$ lower bound, we will then distinguish four regions.
All plots will be cut at $M_1 = 10^{12}\,{\rm GeV}$,
since above this value, according to the condition (\ref{unflavored}),
the unflavored regime is recovered and the asymmetry production
has to switch off.
On the other hand, when the condition Eq. (\ref{full}) is satisfied,
one can expect the fully flavored regime to hold. There is an intermediate
regime where a transition between the fully flavored regime
and the unflavored regime takes place.
This regime will be indicated in all plots with a squared region.
This signals that, even though we still
show the results obtained in the fully flavored regime, important
corrections are expected, especially when $M_1$ gets close to
$\sim 10^{12}\,{\rm GeV}$. Since this region describes
a transition toward the unflavored regime, where the asymmetry
production has to switch off, these corrections
are expected to reduce the final asymmetry,
making more stringent the lower bounds shown in the plots.
Furthermore, since within current calculation, large corrections
to the condition Eq. (\ref{full}) cannot be excluded,
we will also indicate, with a hatched region, that area where
 the condition Eq. (\ref{full}) holds but a very conservative condition,
\be\label{conservative}
M_1\lesssim {10^{11}\,{\rm GeV}\over W_1^{\rm ID}(z_{B}(K_{1\a}))} \,
\ee
does not.
In this region some corrections to the presented
results cannot be excluded but the fully flavored regime should
represent a good approximation.

We anticipate that, in the $N_1$-dominated scenario,
successful leptogenesis always requires $M_1\gtrsim 10^9\,{\rm GeV}$,
where the two-flavor regime holds. Therefore, considering
that we are assuming $\ve_1=\ve_{1\t}+\ve_{1,e+\m}=0$,
the Eq. (\ref{N1DS}) can be specialized into
\be\label{final}
\left. N_{B-L}^{\rm f} \right|_{N_1}
\simeq (\k_{1\,\t}^{\rm f}-\k_{1,e+\m}^{\rm f})\,\ve_{1\t} \, ,
\ee
showing that, in order to have a non-vanishing final asymmetry
it has to be $P_{1\t}^0 \neq P_{1,e+\m}^0$.
The tree-level projectors can be expressed, through
the orthogonal parametrization Eq. (\ref{Opar}), like
\be\label{P01a}
P^0_{1\a}={|\sum_j\,\sqrt{m_j}\,U_{\a j}\,\O_{j 1}|^2
\over \sum_j\,m_j\,|\O^2_{j1}|} \, ,
\ee
that, from the Eq. (\ref{Kimi}), also implies
\be\label{K1alpha}
K_{1\a}=
\left|\sum_j\,\sqrt{m_j\over m_{\star}}\,U_{\a j}\,\O_{j 1}\right|^2 \, .
\ee
Let us now calculate the flavored $C\!P$ asymmetry $\ve_{1\t}$
from the general expression Eq. (\ref{ve1a}). In terms of the
orthogonal parametrization Eq. (\ref{Opar}), this can be re-cast as
\cite{abada2}
\be\label{e1alOm}
r_{1\a}=-\,\sum_{h,l}\,
{m_l\,\sqrt{m_l\,m_h}\over \mt\,m_{\rm atm}}
\,{\rm Im}[U_{\a h}\,U_{\a l}^{\star}\,\O_{h1}\,\O_{l1}]
\, ,
\ee
where we defined $ r_{i\a}\equiv {\ve_{i\a}/ \overline{\ve}(M_i)}$, with
\be
\bar{\ve}(M_i)\equiv {3\over 16\pi}\,{M_i\,m_{\rm atm} \over v^2} \, .
\ee
For real $\O$, the Eq. (\ref{e1alOm}) gets specialized into \cite{abada2}
\be
r_{1\a}=-\sum_{h< l}\,
{\sqrt{m_l\,m_h}\,(m_l-m_h)\over \mt\,m_{\rm atm}}\,
\O_{h1}\,\O_{l1}\,{\rm Im}[U_{\a h}\,U_{\a l}^{\star}] \, .
\ee
Taking $\a=\t$ and specifying  the matrix elements $U_{\a j}$,
from the Eq. (\ref{Umatrix}), one has
\be\label{r1tau}
r_{1\t}=-{m_{\rm atm}\over \mt}\,[A_{12}+A_{13}+A_{23}],
\ee
where
\bea\nonumber
A_{12} & = & - {\sqrt{m_1\,m_2}\,(m_2-m_1) \over m_{\rm atm}^2}
\,\O_{11}\,\O_{21}\,
{\rm Im}[(s_{12}\,s_{23}-c_{12}\,c_{23}\,s_{13}\,e^{i\,\d}) \\ \nonumber
& & \;\;\;\;\;\;\;\;\;\;\;\;\;\;\;\;\;\;\;\;\;\;\;\;
\times \,(c_{12}\,s_{23}+s_{12}\,c_{23}\,s_{13}\,e^{-i\,\d})\,
e^{-{i\over 2}\,(\Phi_2-\Phi_1)}]\, , \\ \nonumber
A_{13} & = & {\sqrt{m_1\,m_3}\,(m_3-m_1)\over m_{\rm atm}^2}
\,\O_{11}\,\O_{31}\,c_{23}\,c_{13}\,
{\rm Im}[(s_{12}\,s_{23}-c_{12}\,c_{23}\,s_{13}\,e^{i\,\d})
e^{{i\over 2}\,\Phi_1}] \, ,
\\ \nonumber
A_{23} & = & - {\sqrt{m_2\,m_3}\,(m_3-m_2) \over m_{\rm atm}^2}
\,\O_{21}\,\O_{31}\,c_{23}\,c_{13}\,
{\rm Im}[(c_{12}\,s_{23}+s_{12}\,c_{23}\,s_{13}\,e^{i\,\d})
\,e^{{i\over 2}\,\Phi_2}]\, .
\eea
In the case of $\d$-leptogenesis ($\Phi_1=\Phi_2=0$)
these expressions further specialize into
\bea\nonumber
A_{12} & = & {\sqrt{m_1\,m_2}\,(m_2-m_1)\over m_{\rm atm}^2}
\,\O_{11}\,\O_{21}\,s_{23}\,c_{23}\,\D
\\ \nonumber
A_{13} & = & -{\sqrt{m_1\,m_3}\,(m_3-m_1)\over m_{\rm atm}^2}
\,\O_{11}\,\O_{31}\,c_{23}^2\,c_{12}\,c_{13}\,\D  \, ,
\\ \nonumber
A_{23} & = & - {\sqrt{m_2\,m_3}\,(m_3-m_2)\over m_{\rm atm}^2}
\,\O_{21}\,\O_{31}\,c_{23}^2\,s_{12}\,c_{13}\,\D \, ,
\eea
where remember that $\D\equiv \sin\theta_{13}\,\sin\d$.

It is now instructive to make some general considerations.
Looking at the expression Eq. (\ref{final}), one can see that,
in order for the final $B-L$ asymmetry not to vanish, two
conditions have to be simultaneously satisfied :
$\ve_{1\t}\neq 0$ and $\k_{1\t}^{\rm f} \neq \k_{1,e+\m}^{\rm f}$.
These two conditions are a specialization of the Sakharov
necessary conditions to the case of $\d$-leptogenesis.
Indeed, the first is the condition to have
$C\!P$ violation and, as one could expect, from the
expressions found for the terms $A_{ij}$, one can have
$\ve_{1\t}\neq 0$ only if $\D\neq 0$.
The second condition is a specialization of the condition of
departure from thermal equilibrium in quite a non-trivial way.
Indeed, in the case of $\d$ leptogenesis, in a full out-of-equilibrium
situation where only decays are active, no final asymmetry is generated
since $\ve_1=0$, implying that there is an equal number
of decays into lepton and anti leptons. However, the presence of inverse
processes can remove this balance, yielding a different wash-out rate
for the $\t$ asymmetry
and for the $e+\m$ asymmetry, such that, if $K_{1\t}\neq K_{1,e+\m}$,
one has a net lepton number dynamical generation. From the expression
(\ref{K1alpha}), one can see that this is possible independently of the
value of the Dirac phase that, therefore, is directly responsible only
for $C\!P$ violation and not for lepton number violation,
exactly as in neutrino mixing, where indeed lepton number is conserved.
It should also be noticed that the $\ve_{1\a}$'s  are expressed through
quantities ${\rm Im}[U_{\a h}\,U^{\star}_{\a l}]$, that are invariant
under change of the PMNS matrix parametrization \cite{nieves,pascoli2}.
Therefore, the final asymmetry depends correctly
only on physical quantities.

Maximizing the asymmetry over all involved parameters
for fixed $M_1$ and $K_1$ and
imposing $\eta_B^{\rm max} \geq \eta_B^{\rm CMB}$ (cf. (\ref{etaB}) and
(\ref{etaBobs})), a lower bound on $M_1$ is obtained \cite{flavorlep}
\be
M_1 \geq M_1^{\rm min}(K_1)\equiv {\overline{M}_1\,\over
\k^{\rm f}_1(K_1)\,\xi_1^{\rm max}} \, ,
\ee
where we introduced the quantity
\footnote{Notice that $\overline{M_1}$ gives the
lower bound on $M_1$ in the unflavored case
for initial thermal abundance and in the limit $K_1\rightarrow 0$.
Because of the improved determination of $\eta_B^{\rm CMB}/m_{\rm atm}$
from the 3 years WMAP data \cite{WMAP3} and from new data from
neutrino oscillation experiments, in particular from the MINOS experiment,
the error on $\overline{M}_1$ is halved compared to the previous estimation
in \cite{annals}.}
\be\label{barM1}
\overline{M}_1 \equiv {16\,\pi\over 3}\,
{N_{\g}^{\rm rec}\,v^2\over a_{\rm sph}}\,
{\eta_B^{\rm CMB}\over m_{\rm atm}}
=(6.25\pm 0.4)\times 10^8\,{\rm GeV} \gtrsim 5\times 10^8{\rm GeV} \, .
\ee
The last inequality gives the $3\s$  value that we
used to obtain all the results shown in the figures.
We also defined \cite{flavorlep}
\be\label{xi1a}
\xi_1 \equiv \sum_{\a=\t,e+\m}\,\xi_{1\a}\, ,
\hspace{10mm}\mbox{with}\hspace{12mm}
\xi_{1\a}\equiv {r_{1\a}\,\k_{1\a}^{\rm f}(K_{1\a})
                      \over \k_1^{\rm f}(K_1)}\, ,
\ee
that gives the deviation introduced by flavor
effects compared to the unflavored approximation in the
hierarchical light neutrino case ($m_1=0$).
Notice that $r_{1\t}\propto \D$, implying $N_{B-L}^{\rm f}\propto \D$ as well.
Therefore, the maximum asymmetry is obtained for $|\d|=\pi/2$
and $s_{13}=0.20$.


The calculation of the contribution to the asymmetry from $N_2$-decays
proceeds in an analogous way. Again this can always be calculated in the
two-flavor regime, since, in the HL, successful leptogenesis always implies
$M_2\gtrsim 10^{9}\,{\rm GeV}$.
Therefore, one can write an expression similar to the Eq. (\ref{final})
for the contribution to the final asymmetry from $N_2$-decays,
\be\label{final2}
\left.N_{B-L}^{\rm f}\right|_{N_2}\simeq
(\k_{2\,\t}^{\rm f}-\k_{2,e+\m}^{\rm f})\,\ve_{2\t} \, .
\ee
The difference is now in the calculation of the efficiency factors
that are suppressed by the wash-out of the $N_1$ inverse
processes. In the HL this additional wash-out factorizes and
\cite{beyond,flavorlep,vives}
\be
\k_{2\alpha}^{\rm f}\simeq \k(K_{2\alpha})\,
e^{-{3\,\pi\over 8}\,K_{1\a}}\, ,
\ee
where $K_{2\a}\equiv P_{2\a}^0\,K_2$. For the calculation of the
tree-level projectors $P_{2\a}^0$ an expression
analogous to the Eq. (\ref{P01a}) holds.


The calculation of the contribution to the final asymmetry
from $N_3$-decays proceeds in a similar way and analogous
expressions hold. The only non trivial difference is that
now, in the calculation of the efficiency factors,
one has also to include the wash-out from the $N_2$ inverse processes,
so that
\be\label{k3a}
\k_{3\alpha}^{\rm f}\simeq \k(K_{3\alpha})\,
e^{-{3\,\pi\over 8}\,(K_{1\a}+K_{2\a})}\, .
\ee
Notice that in the calculation of $\k_{2\a}^{\rm f}$
($\k_{3\a}^{\rm f}$) we are not including a possible effect where
part of the
asymmetry in the flavor $\a=e+\m$ produced in $N_2$ ($N_3$-decays) is
orthogonal to $N_1$ inverse decays \cite{bcst,nardi2} and is
not washed out. This
wash-out avoidance does not apply to the asymmetry in the
$\t$ flavor. Therefore, as we have verified, in all cases we
have considered the effect is negligible, since a $\t$-dominated
scenario is always realized.


Let us now calculate the final asymmetry in some interesting cases.

\subsection{$\O=R_{13}$}

The first case we consider is $\O=R_{13}$, implying $A_{12}=A_{23}=0$
in the Eq. (\ref{r1tau}). As we said already, it is easy to check from the
Eq. (\ref{ve2a}) that $\ve_{2\t}=0$ and therefore there is no
asymmetry production from $N_2$-decays even if $M_2\lesssim 10^{12}\,{\rm GeV}$.
On the other hand, one obtains
\be
r_{3\tau}=
-{2\over 3}\,{\sqrt{m_1\,m_3}\,(m_3-m_1) \over \mttt\,m_{\rm atm}}\,
\o_{31}\,\sqrt{1-\o_{31}^2}\,c_{12}\,c_{23}^2\,c_{13}\,\D \, ,
\ee
essentially the same expression as
for $r_{1\t}$ but with $\mt$ replaced by $\mttt$.
Therefore, for $M_3\lesssim 10^{12}\,{\rm GeV}$,
one has to worry about a potential non-negligible contribution
from $N_3$ decays. However, when the wash-out
from $N_1$ and $N_2$ inverse processes is taken into account,
see Eq. (\ref{k3a}), we always find that the
contribution from $N_3$-decays is negligible and
the $N_1$-dominated scenario holds.

The results are shown in Fig. 1 for $s_{13}=0.20$, $\d=-\pi/2$
and $m_1/m_{\rm atm}=0.1$, a choice of values that approximately
maximizes the final asymmetry and yields the lower bound $M_1^{\rm min}(K_1)$.
\begin{figure}
\hspace*{-5mm}
\psfig{file=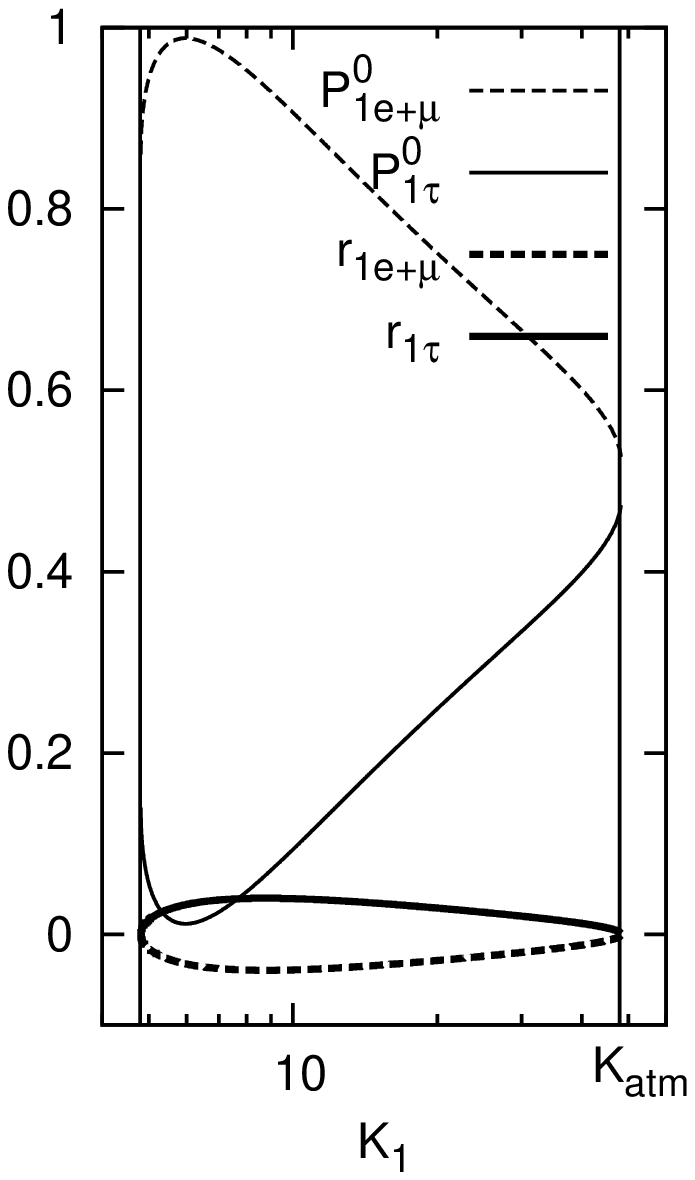,height=7cm,width=53mm}
\hspace{-1mm}
\psfig{file=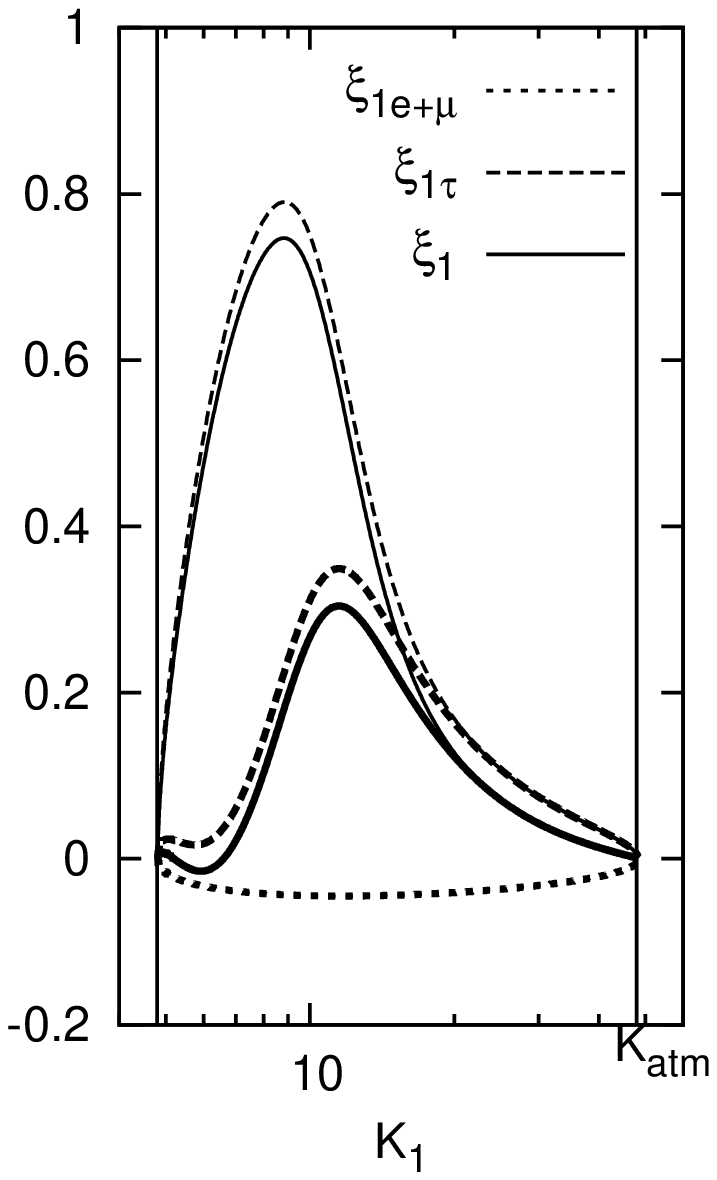,height=7cm,width=53mm}
\hspace{-1mm}
\psfig{file=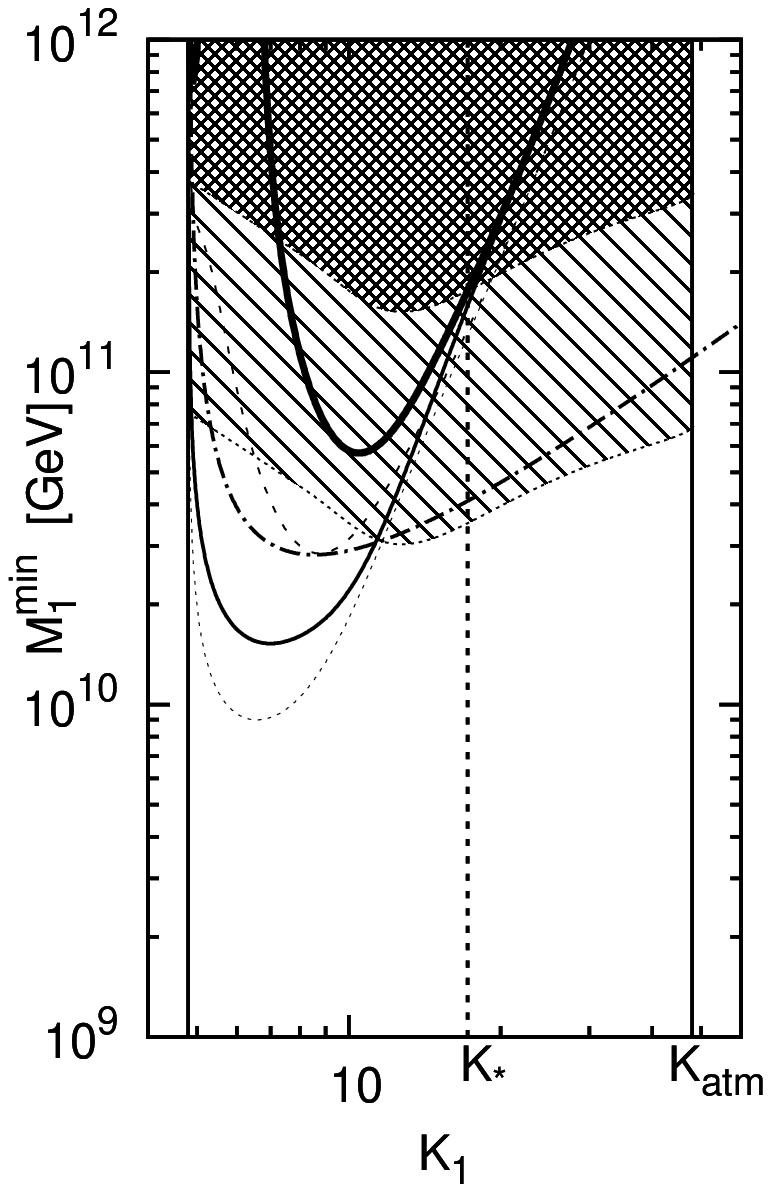,height=7cm,width=53mm}
\caption{Dependence of different quantities on $K_1$ for
$m_1/m_{\rm atm}=0.1$, $s_{13}=0.2$, $\d=-\pi/2$ and real $\O=R_{13}$
with $\o_{31}<0$.
Left panel: projectors $P^0_{1\a}$ and $r_{1\a}$;
central panel: $\xi_{1\a}$ and $\xi_1$ as
defined in Eq. (\ref{xi1a}) for thermal (thin)
and vanishing (thick) initial abundance;
right panel: lower bound on $M_1$ for thermal (thin solid) and
vanishing (thick solid) initial abundance compared with the one-flavor
approximation result (dash-dotted line) obtained for complex $\O=R_{13}$.
In the squared region the condition Eq. (\ref{full}) is not
satisfied and in the hatched region even the more
conservative condition Eq. (\ref{conservative}) is not satisfied.
The dotted lines (thick for vanishing initial abundance and thin
for thermal initial abundance) correspond still to a real
$\O=R_{13}$ but this time $\d=0$ while the only non vanishing low energy
phase is the Majorana phase $\Phi_1=-\pi/2$.}
\end{figure}
In the left panel we show the tree level projectors
$P^0_{1\a}$ and the $r_{1\a}$'s.
It can be seen how for $K_1\gg 10$ one has
$P^0_{1\t}\simeq P^0_{1,e+\m}\simeq  1/2$,
while for $K_1\sim 10$ one has $P^0_{1\t}\ll P^0_{1,e+\m}$.
In the central panel $\xi_1$ and the $\xi_{1\a}$'s  are plotted and one can
see how for $K_1\simeq 10$ a $\t$-dominance is realized. Finally, in the
right panel, we show $M_1^{\rm min}(K_1)$ and we compare it with the
lower bound in the unflavored approximation obtained for
$\O=R_{13}$ (in this case $\O$ cannot be real) \cite{geometry}.
One can see how, at $K_1\gg 10$, the asymmetry production
rapidly dies, so that $\xi_1\ra 0$ and $M_1^{\rm min}(K_1)\ra \infty$.
 Notice that we plotted the lower bound both
for initial thermal $N_1$-abundance and for
initial vanishing $N_1$-abundance. We also indicated $K_{\star}$,
defined as that value of $K_1$ such that for $K_1\gtrsim K_{\star}$
the dependence on the
initial conditions can be neglected and the strong wash-out regime
holds. One can notice that the intermediate regime between a fully flavored
regime and the unflavored regime, the squared area, is quite extended.
In this regime corrections to the results we are showing, obtained in the
 fully flavored regime, are expected, in a way that the unflavored regime
should be recovered for $M_1\rightarrow 10^{12}\,{\rm GeV}$. In this limit
the asymmetry production has to switch off and therefore one expects
that the lower bound on $M_1$ has to become more restrictive
and eventually, for $M_1\rightarrow 10^{12}\,{\rm GeV}$,
the allowed region has to close up. Therefore,
one can see that there is no allowed region in the strong wash-out regime.
The hatched area, where corrections cannot be excluded within current
theoretical uncertainties, cuts away almost completely any allowed region
even in the weak wash-out regime. In conclusion, the allowed
region where one can safely rely on the fully flavored regime
according to current calculations, is very restricted and confined
only to a small region in the weak wash-out regime.


\subsection{$M_3\gg 10^{14}\,{\rm GeV}$}

The second case we consider is the limit $M_3\gg 10^{14}\,{\rm GeV}$.
 In this limit one has necessarily $m_1\ll m_{\rm sol}$,
implying $m_3\simeq m_{\rm atm}$, and also \cite{fgy,ir,ct}
\be\label{ss}
\O=
\left(
\begin{array}{ccc}
     0            &   0                 &  1 \\
\sqrt{1-\O^2_{31}}& -\O_{31}            &  0 \\
       \O_{31}    & \sqrt{1-\O^2_{31}}  &  0
\end{array}
\right)
\, .
\ee
Notice that this particular form of $\O$ corresponds to set
$\o_{32}=1$ and $\o_{21}=1$ in the Eq. (\ref{R}).
Now in the expression for $r_{1\t}$ (cf. Eq. (\ref{r1tau}))
one has $A_{12}=A_{13}=0$ and therefore
\be
r_{1\t}\simeq
{m_{\rm atm}\over \mt}\,
\sqrt{m_2 \over m_{\rm atm}}\,
\left(1-{m_2\over m_{\rm atm}}\right)\,
\,\O_{31}\,\sqrt{1-\O^2_{31}}\,c_{23}^2\,c_{13}\,s_{12}\,\D \, .
\ee
If  $M_2\gtrsim 10^{12}\,{\rm GeV}$,
there is no contribution from the next-to-lightest RH neutrino decays
anyway, since these occur in the unflavored regime where $\ve_2\simeq 0$.
On the other hand, if $M_2\lesssim 10^{12}\,{\rm GeV}$, then one has
to worry about a (flavored) asymmetry generation from $N_2$-decays.
A calculation of $\ve_{2\a}$ shows that
the first term in the Eq. (\ref{ve2a}) vanishes while the second term gives
\be
r_{2\t}=-{2\over 3}\,{m_{\rm atm}\over \mtt}\,
              \sqrt{m_2\over m_{\rm atm}}\,
               \left(1-{m_2\over m_{\rm atm}}\right)\,\O_{31}\,
               \sqrt{1-\O^2_{31}}\,c_{23}^2\,s_{12}\,\D \,.
\ee
This is an example of how the second term in the Eq. (\ref{ve2a})
is not suppressed in the HL like the first term.
However, like for the contribution from $N_3$-decays
in the case $\O=R_{13}$, when the wash-out from $N_1$-inverse processes
is taken into account one finds
$\left. N_{B-L}^{\rm f}\right|_{N_2}\ll \left.N_{B-L}^{\rm f}\right|_{N_1}$
and a $N_1$-dominated scenario is realized anyway.

Notice that there is a strong dependence whether one assumes
a normal or an inverted hierarchy. For normal hierarchy
the results are shown in Fig. 2 for $\o_{31}>0$ and $\d=\pi/2$.
\begin{figure}
\hspace*{-5mm}
\psfig{file=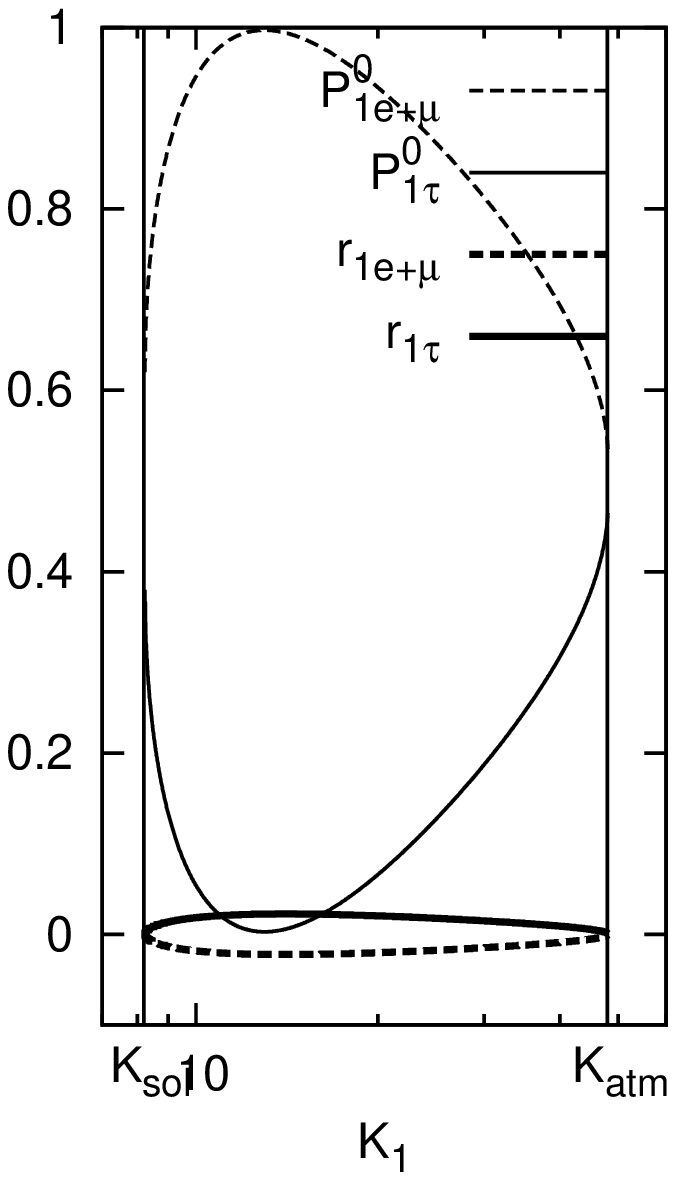,height=7cm,width=53mm}
\hspace{-1mm}
\psfig{file=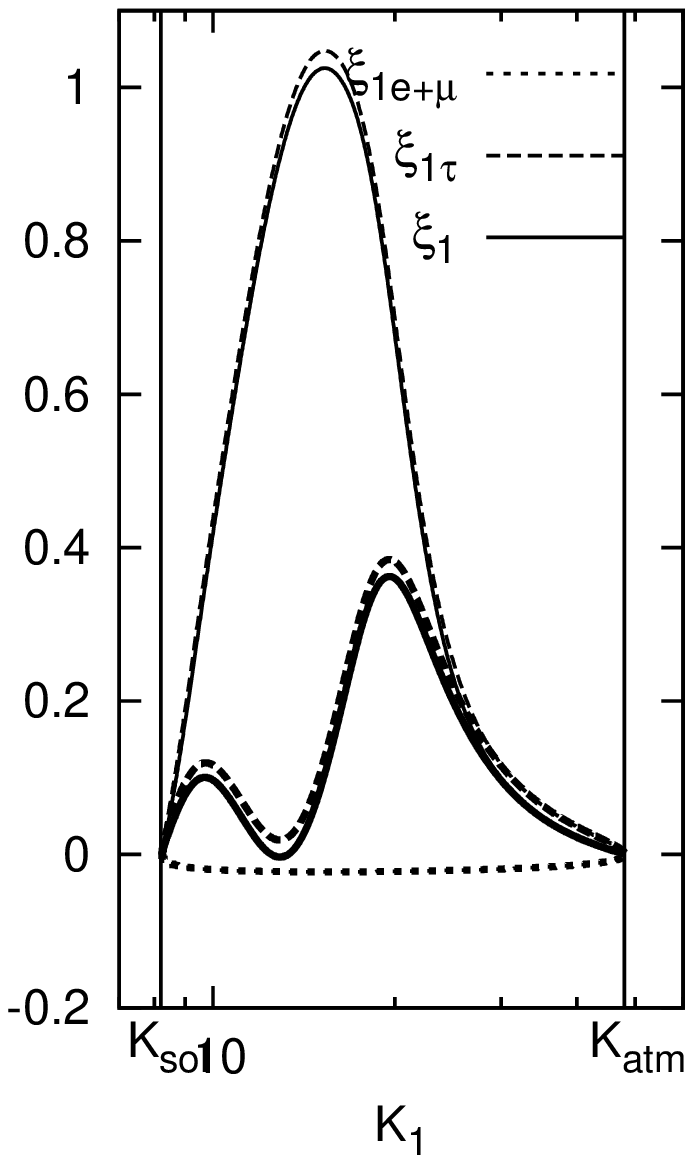,height=7cm,width=53mm}
\hspace{-1mm}
\psfig{file=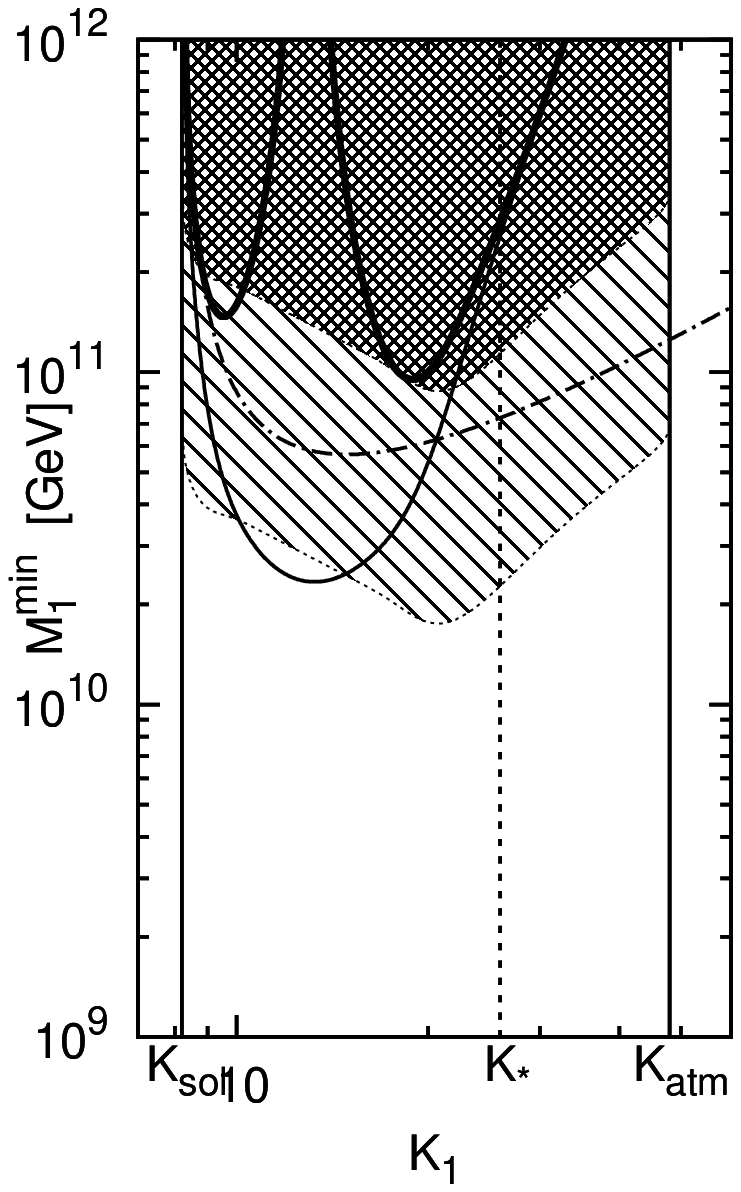,height=7cm,width=53mm}
\caption{Plots as in Fig. 1 but for the case
$M_3\gg 10^{14}\,{\rm GeV}$, corresponding to the special
form of $\O$ in the Eq. (\ref{ss}). Here we are moreover
assuming $M_2\gtrsim 10^{12}\,{\rm GeV}$, normal hierarchy.
The lower bound $M_1^{\rm min}(K_1)$ is obtained for
$\o_{31}>0$ and $\d=\pi/2$.}
\end{figure}
For inverted hierarchy the asymmetry is so suppressed that
there is no allowed region. This means that for any choice of
the parameters one always obtains
$M_1^{\rm min}\gtrsim 10^{12}\,{\rm GeV}$.

Notice that results for $\d$-leptogenesis, in this particular
case where $M_3\gg 10^{14}\,{\rm GeV}$, have been recently
presented in \cite{pascoli2} for vanishing initial $N_1$ abundance.
For example in \cite{pascoli2} the authors obtain a lower bound
$\sin\theta_{13}\gtrsim 0.09$ imposing the existence of an allowed region
for $M_1\lesssim 5\times 10^{11}\,{\rm GeV}$ while
we would obtain $\sin\theta_{13}\gtrsim 0.05$.
The difference is probably due to a ($\sim 30\%$)
more conservative lower bound that we are using on $\overline{M}_1$
(see Eq. (\ref{barM1})), a difference in the employed value
of $m_{\star}$ (see Eq. (\ref{d})), only partly understood in terms
of the different convention for the Higgs v.e.v $v$. There is also
a difference in the employed efficiency factor in the strong wash-out regime
that, in our case, is about a factor 2 larger.
Another likely minor source of difference
is that we are not accounting for the effect of spectator processes
encoded in the matrix $A$ that relates
the $B/3-L_{\alpha}$ asymmetries to the $L_{\alpha}$ asymmetries \cite{bcst}.
However, notice that here we do not want to emphasize too much
a precise value of this lower bound on $\sin\theta_{13}$, since we believe
this is anyway affected by much larger theoretical uncertainties on the validity
of the fully flavored regime. It is however a good way to compare our results
with those presented in \cite{pascoli2}.


\subsection{$\O=R_{12}$}

The third case we consider is $\O=R_{12}$.
This time one has $A_{13}=A_{23}=0$ in the Eq.(\ref{r1tau}).
In the case of normal hierarchy the $C\!P$ asymmetry, compared
to the case $\O=R_{13}$, is
suppressed by a factor $(m_{\rm sol}/m_{\rm atm})^{3/2}$,
while it is essentially the same for inverted hierarchy.
The projectors present very similar features to the case $\O=R_{13}$.
One can also again calculate, for
$M_2\lesssim 10^{12}\,{\rm GeV}$, the contribution from $N_2$-decays
to the final asymmetry and again one finds that
the first term in the Eq. (\ref{ve2a})
vanishes, while the second produces a term $\propto M_1$, so that
\be
r_{2\t}=
{2\over 3}\,{\sqrt{m_1\,m_2}\,(m_2-m_1) \over \mtt\,m_{\rm atm}}\,
\o_{21}\,\sqrt{1-\o_{21}^2}\,s_{23}\,c_{23}\,\D \, .
\ee
When the efficiency factors are taken into account, one finds that
only in the case of normal hierarchy the contribution to
the final asymmetry from $N_2$-decays can be comparable
to that one from $N_1$-decays. However, in this case
both productions are suppressed and there
is no allowed region anyway in the end.
In the case of inverted hierarchy, the contribution
from $N_2$-decays is always negligible compared to that one
from $N_1$-decays. Notice, moreover, that
$\ve_{3\a}=0$ and therefore  there is no contribution from
$N_3$-decays.
In conclusion, for $\O=R_{12}$, the lower bound on $M_1$ for normal hierarchy
is much more restrictive than in the case $\O=R_{13}$, while it is very similar
for inverted hierarchy. A production from the two heavier RH
neutrinos can be neglected and the $N_1$-dominated scenario always
holds when the asymmetry is maximized.


\subsection{$\O=R_{23}$}

The last interesting case is $\O=R_{23}$.
From the Eq. (\ref{e1alOm}) one can easily check
that $\ve_{1\a}=0$. One can also easily check that, contrarily to the
case $\O=R_{12}$, the second term in the Eq. (\ref{ve2a}) vanishes while
the first term does not and yields
\be
r'_{2\t} \equiv {\ve_{2\t}\over \bar{\ve}(M_2)} =
{\sqrt{m_2\,m_3}\,(m_3-m_2)\over \mtt\,m_{\rm atm}}\,
\o_{32}\,\sqrt{1-\o_{32}^2}\,s_{12}\,c_{23}^2\,c_{13}\,\D \, .
\ee
Notice that this time $\ve_{2\t}\propto M_2$ and actually, more generally,
one can see that this expression is obtained from the Eq. (\ref{r1tau})
for $r_{1\t}$ in the case $\O=R_{13}$, just with the replacement
$(M_1,\mt)\rightarrow (M_2,\mtt)$.
At the same time one has $K_1=m_1/m_{\star}$ so that
the wash-out from $N_1$-inverse processes vanishes for $m_1\rightarrow 0$.
For $M_3\lesssim 10^{12}\,{\rm GeV}$ one has to worry about
a possible contribution to the asymmetry also from $N_3$-decays.
A straightforward  calculation shows that
$\ve_{3\a}=(2/3)\ve_{2\a}$ and therefore an asymmetry is
produced at $T\sim M_3$. However, we verified, once more, that
the wash-out from $N_2$-inverse processes is always strong enough
that the contribution to the final asymmetry
from $N_3$-decays is negligible.

In complete analogy with the unflavored
case \cite{geometry}, one has that the lower bound $M_1^{\rm min}(K_1)$
is replaced by a lower bound $M_2^{\rm min}(K_2)$ obtained
for $\o_{32}>0$ and shown in the right panel of Fig. 3.
\begin{figure}
\hspace*{-5mm}
\psfig{file=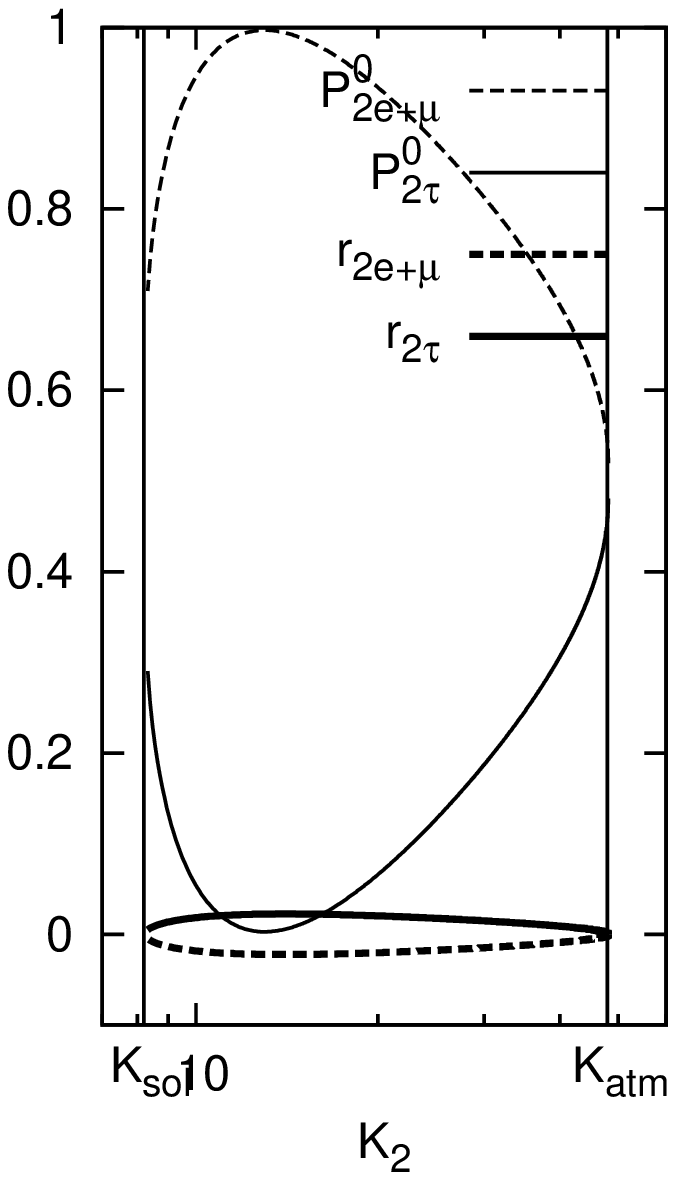,height=7cm,width=53mm}
\hspace{-1mm}
\psfig{file=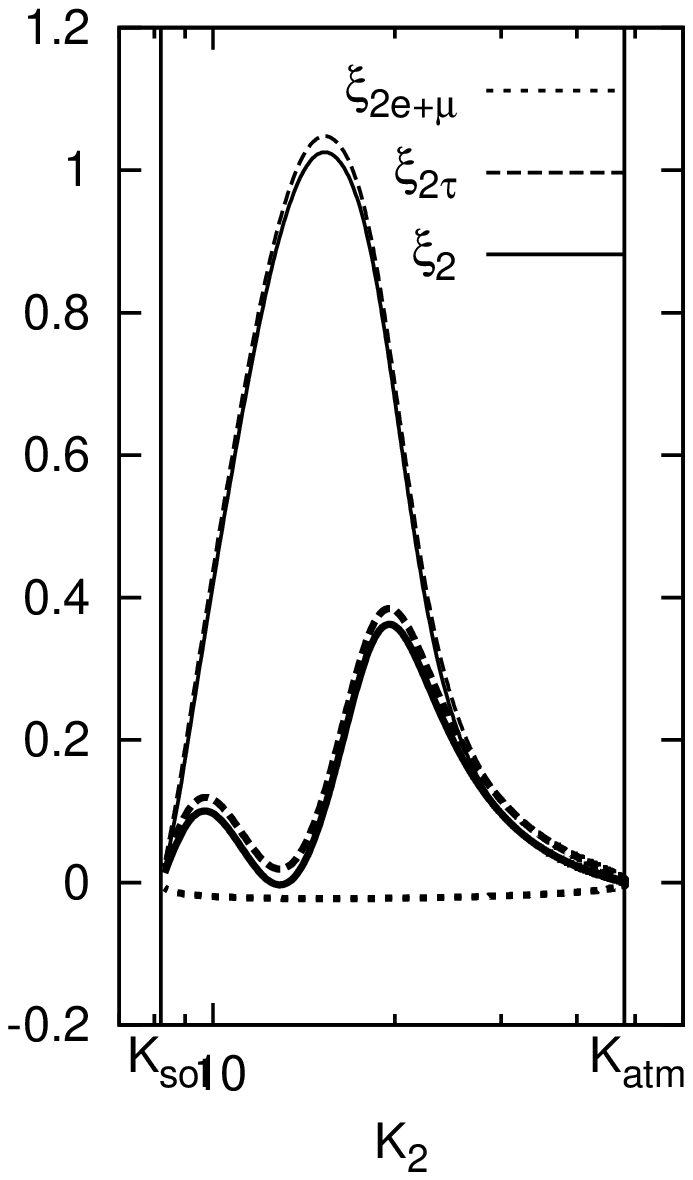,height=7cm,width=53mm}
\hspace{-1mm}
\psfig{file=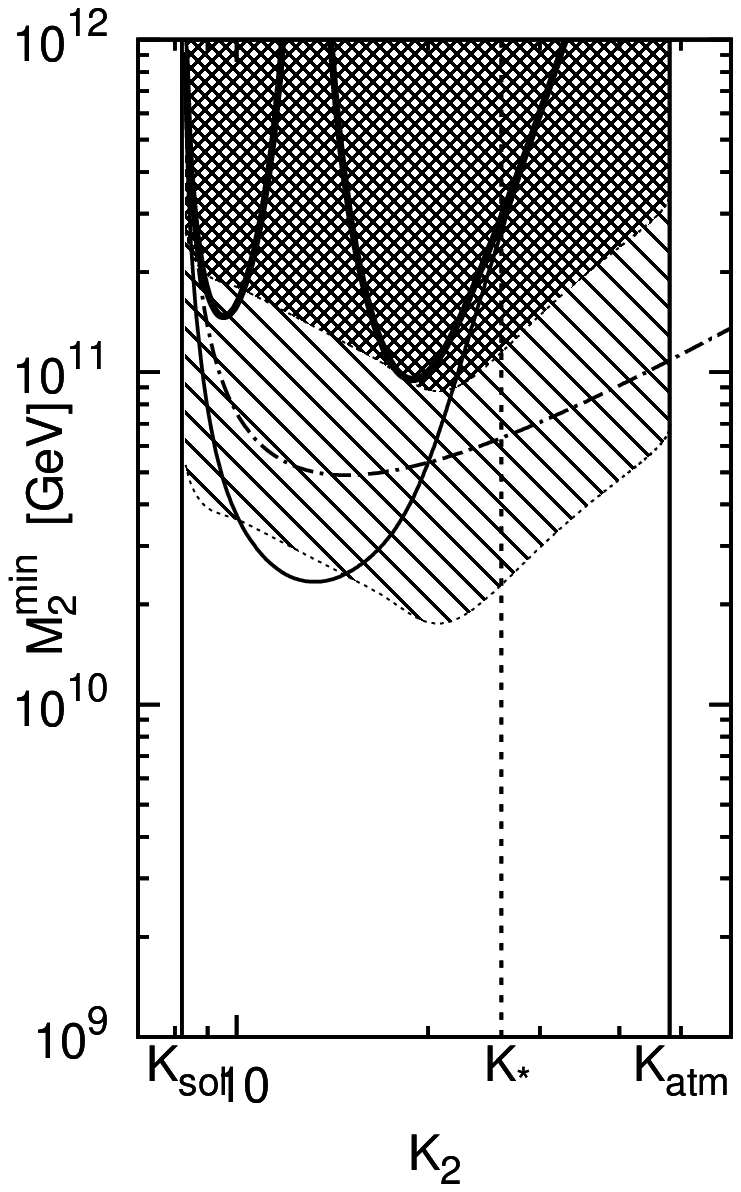,height=7cm,width=53mm}
\caption{Dependence of different quantities on $K_2$ for
$m_1=0$, $s_{13}=0.2$, $\d=\pi/2$ and real $\O=R_{23}$ with $\o_{32}>0$.
Left panel: projectors $P^0_{2\a}$ and quantities $r'_{2\a}$;
central panel: $\xi_{2\a}$ and $\xi_2$ for thermal (thin)
and vanishing (thick) initial abundance;
right panel: lower bound on $M_2$ for thermal (thin solid) and
vanishing (thick solid) abundance compared with the one-flavor
approximation result (dash-dotted line) as obtained in \cite{geometry}.}
\end{figure}
One can see that also in this case, within
the validity of the condition Eq. (\ref{full}),
the allowed region is constrained to a small portion
falling in the weak wash-out regime.
Assuming the very conservative condition of validity for
the fully flavored regime, outside the squared and hatched
regions, there is no allowed region even in the weak wash-out regime.


One can wonder whether there is some choice of $\O$,
beyond the special cases we analyzed, where the final asymmetry
is much higher and the lower bound on $M_1$ much more
relaxed, especially in the strong wash-out regime. We have checked
different intermediate cases and we can exclude such a possibility.
Therefore, the lower bound shown in Fig. 1 has to be considered, with good
approximation, the lowest bound for any choice of real $\O$.

Another legitimate doubt is whether, going beyond the approximations we made,
the lower bound in Fig. 1 can be considerably relaxed.
However, the inclusion of non resonant
$\D L=2$ or $\D L=1$ scattering does not produce large corrections.
Recently the effect of the off-diagonal terms in the $A$ matrix
has been considered, but it
has been shown that it does not produce any relevant change
in the final asymmetry \cite{abada3}.

Relevant corrections, as already pointed out,
can come only from a full quantum kinetic treatment, that should
describe accurately the transition between the unflavored regime
and the fully flavored regime.

The same kind of considerations holds for the $N_2$-dominated scenario,
realized for $\O=R_{23}$. As soon as $\O$ deviates
from $R_{23}$, the wash-out from $N_1$ inverse processes comes into play
suppressing the final asymmetry and at the same time $\ve_{2\t}$
gets also suppressed. Therefore, the lower bound on $M_2$ is necessarily
obtained for $\O=R_{23}$ in complete analogy with the unflavored
approximation \cite{geometry}.

In conclusion $\d$-leptogenesis in the HL is severely constrained,
confirming the conclusions of \cite{flavorlep} and \cite{antusch}.
In particular, imposing independence of the initial conditions,
then not even a marginal allowed region seems to survive.
Notice moreover that all plots have been obtained for $s_{13}=0.2$,
the current $3\,\s$ upper bound value. Assuming that for
values of $M_1$ above the condition Eq. (\ref{full})  the
unflavored regime is quickly recovered and therefore that
the asymmetry production quickly switches off, then
a one-order-of-magnitude improvement of the upper bound on $\sin\theta_{13}$
would essentially completely rule out $\d$-leptogenesis in the HL,
even the marginally allowed regions falling in the weak wash-out regime.

Therefore, in the next section, we will consider the effect
of close heavy neutrino masses in enhancing
the $C\!P$ asymmetries and relaxing the lower bounds
on $M_1, M_2$ and the related one on $T_{\rm reh}$.
In the end of this section we want to mention that
in the more general case of real $\O$ with non-vanishing Majorana phases,
an upper bound $m_1\lesssim 0.1\,{\rm ev}$
has been obtained in the fully flavored regime \cite{branco}.
This bound clearly applies also to $\d$-leptogenesis, but in this case,
considering the results we have obtained and  the expected quantum
kinetic corrections to the fully flavored regime,
the issue is actually whether an allowed region exists at all in the HL,
even for $m_1=0$. Therefore, we do not even try to place an upper bound
on $m_1$ in the HL. In the next section, we will show
that actually for $\d$-leptogenesis an upper bound on $m_1$
holds even in the resonant limit, where the $C\!P$ asymmetries are
maximally enhanced.

\section{The degenerate limit}

In this section we show that going beyond the HL the lower
bound on $M_1$ (or on $M_2$) can be considerably relaxed.
Nevertheless, we will see that some interesting constraints
on the involved parameters still apply.
For simplicity, we can assume a full three-flavor regime holding
for $M_1$ (or $M_2$) $\ll 10^{9}\,{\rm GeV}$, when also the muon-Yukawa
interactions are faster than inverse decays.
Therefore now, when we sum over the flavor index,
it has to be meant  $\a=e,\mu,\tau$. This assumption
simplifies the calculation, since we do not have
to describe a transition between the two and the three-flavor
regime and because we can
completely neglect the effect, envisaged in \cite{bcst,nardi2}, for which
part of the asymmetry produced from $N_2$-decays is not touched
by $N_1$-inverse decays.
Indeed in a two-flavor regime, even though in the HL we have found that this
effect is negligible in all cases we considered because a
$\t$-dominance is always realized, in the DL
it can become more relevant because the asymmetry
is not necessarily produced dominantly in the $\t$-flavor.

In order to go beyond the HL, it is convenient to introduce the quantities
\be
\delta_{ji}\equiv {M_j-M_i\over M_i} = \sqrt{x_j\over x_i}-1 \, .
\ee
We are interested in the degenerate limit (DL),
where at least one $\d_{ji}$ is small enough that
both the asymmetry production from decays and the wash-out from inverse
processes of the $N_i$'s and of the $N_j$'s can be approximately treated as
if they occur at the same temperature, so that they can be simply added up.
The DL is a good approximation for $|\delta_{ji}|\lesssim 0.01$ \cite{beyond}.
If $i,j\neq 3$ and $M_3\gg M_2\simeq M_1$ then
one has a partial DL and  in this case
the efficiency factors can be approximated,
for thermal initial abundance, as
\cite{beyond}
\be\label{k1}
\k_{i\a}^{\rm f}\simeq \k_{j\a}^{\rm f}\simeq
\k(K_{i\a}+K_{j\a}) \, .
\ee
In all considered cases, it will be always verified
$K_{i\a}+K_{j\a}\gg 1$, so that the strong wash-out regime always
applies and there is no need to consider the case of initial
vanishing abundance. Another possibility
is to have a partial DL with $i,j\neq 1$ so that $M_1\ll M_2\simeq M_3$.
In this case one has to take into account the wash-out
from the lightest RH neutrino and therefore
\be\label{k2}
\k_{i\a}^{\rm f}\simeq \k_{j\a}^{\rm f}\simeq
\k(K_{i\a}+K_{j\a})\,e^{-{3\pi\over 8}K_{1\a}} \, .
\ee
Finally, in the full DL, one has $M_1\simeq M_2 \simeq M_3$ and
\be\label{k3}
\k_{1\a}^{\rm f}\simeq \k_{2\a}^{\rm f}\simeq \k_{3\a}^{\rm f}
\simeq \k(K_{1\a}+K_{2\a}+K_{3\a}) \, .
\ee
Let us now calculate the flavored $C\!P$ asymmetries.
In the case of real $\O$,
implying real $(h^{\dagger}\,h)_{ij}=(h^{\dagger}h)_{ji}$,
the general expression Eq.'s (\ref{veia}) can be conveniently specialized as
\be\label{veiarealO}
\ve_{i\a}=
\frac{3}{16 \p (h^{\dag}h)_{ii}}
\sum_{j\neq i}\,(h^{\dag}h)_{i j}\,{\rm Im}
\left[h_{\a i}^{\star}\,h_{\a j}\right]\,
\left[\frac{\x(x_j/x_i)}{\sqrt{x_j/x_i}}+
\frac{2}{3(x_j/x_i-1)}\right] \, .
\ee
In the DL one has approximately $\xi(x_j/x_i)\simeq 1/(3\,\d_{ji})$
and consequently
\be
\ve_{i\a}\simeq
\frac{1}{8\,\p (h^{\dag}h)_{ii}}
\sum_{j\neq i}\,(h^{\dag}h)_{i j}\,
{\rm Im}\left[h_{\a i}^{\star}\,h_{\a j}\right]\,\d_{ji}^{-1}.
\ee
We can again express the neutrino Yukawa coupling matrix through the
orthogonal representation. This time the
presence of the factor $\d_{ji}^{-1}$ does not allow to remove the sum on $j$,
as it has been possible in the HL in order to derive the Eq. (\ref{e1alOm}).
However, considering the same special cases studied  in the HL,
only one term $j\neq i$ survives and we can write
\be
\ve_{i\a}\simeq {2\,\bar{\ve}(M_i)\over 3\,\delta_{ji}}\,
\sum_{n,h<l}\,{m_n\,\sqrt{m_h\,m_l}\over\mti\,m_{\rm atm}}\,
\O_{ni}\,\O_{nj}
\,[\O_{hi}\,\O_{lj}-\O_{li}\,\O_{hj}]
\,{\rm Im}[U^{\star}_{\a h}\,U_{\a l}] \, .
\ee
The same expression holds for $\ve_{j\alpha}$
simply exchanging the $i$ and $j$ indexes.
We can always choose $j>i$, so that $M_j\geq M_i$.
In all the particular cases we will consider
it is realized $\ve_{k\a}=0$, for $k\neq i,j$, and moreover
the following simplifications apply:
\be
\sum_n\,m_n\,\O_{ni}\,\O_{nj}=(m_q-m_p)\,\O_{ji}\,\O_{jj}
\hspace{5mm}
{\rm and}
\hspace{5mm}
\sum_{h<l}\,{\sqrt{m_h\,m_l}}\,[\O_{hi}\,\O_{lj}-\O_{li}\,\O_{hj}]
=\sqrt{m_q\,m_p} \, ,
\ee
with $q>p$. Except for $M_3\gg 10^{14}\,{\rm GeV}$, in the other cases
one has $q=j$ and $p=i$. The final asymmetry can then be expressed as
\be
N_{B-L}^{\rm f} \simeq
\sum_{\a}\,(\ve_{i\a}+\ve_{j\a})\,
\k_{\a}^{\rm f}(K_{i\a}+K_{j\a},K_{k\a})=
{\bar{\ve}(M_i)\over 3\,\d_{ji}}\,g(m_1,\O_{ji},\theta_{13},\d)\,\D\, ,
\ee
where
\bea \nonumber
g(m_1,\O_{ji},\theta_{13},\d) & \equiv &
{2\,K_{\rm atm}\,(K_i+K_j)\over K_i\,K_j}\,
\,{(m_q-m_p)\,\sqrt{m_q\,m_p}\over m_{\rm atm}^2}\,
\,
\O_{ji}\,\sqrt{1-\O_{ji}^2} \\ &  & \times   \label{g}
\,\sum_{\alpha}\,\k_{\alpha}^{\rm f}(K_{i\a}+K_{j\a},K_{k\a})\,
{{\rm Im}[U^{\star}_{\a p}\,U_{\a q}]\over\D}\,
\eea
and where
$\k_{\a}^{\rm f}(K_{i\a}+K_{j\a},K_{k\a})=\k_{i\a}^{\rm f}=\k_{j\a}^{\rm f}$
is given by one of the three expressions Eq. (\ref{k1}), Eq. (\ref{k2}) or
Eq. (\ref{k3}), according to the particular case.
It is interesting to notice that in the
full DL the expression (\ref{k3}) holds and
as a consequence of the orthogonality of $\O$, one has
\be
K_{1\a}+K_{2\a}+K_{3\a}=\sum_k\,{m_k\over m_{\star}}\,|U_{\a k}|^2 \, .
\ee
In the degenerate limit, since $U$ is unitary,
this quantity tends to $m/m_{\star}$, independently of the flavor, and
therefore the sum on the flavors in the Eq. (\ref{g}) tends to
vanish. This will contribute, as we will see, to place a stringent
upper bound on the absolute neutrino mass scale in the full DL.

It is also worthwhile to notice that the sign of $\D$ cannot be predicted
from the sign of the
observed final asymmetry, since the sign of $g(m_1,\O_{ji},\theta_{13},\d)$
depends on the sign of $\O_{ji}$ that is undetermined. Notice also that
${{\rm Im}[U^{\star}_{\a h}\,U_{\a l}]/\D}$ does not depend on $\D$ but
nevertheless there is a dependence of $g(m_1,\O_{ji},\theta_{13},\d)$
on $\d$ and on $\theta_{13}$ coming from the tree level projectors
$P^0_{i\a}$ in the sum $K_{i\a}+K_{j\a}$.
However, in any case, for $\D\rightarrow 0$
one has $g(m_1,\O_{ji},\theta_{13},\d)\,\D\rightarrow 0$,
since the final asymmetry has to vanish when
$\sin\theta_{13}$ or $\sin\d$ vanish.

The function $|g(m_1,\O_{ji},\theta_{13},\d)|$ can be
maximized over $\O_{ji}$. Indeed for $m_1=0$, since $\k< 1$ and
$K_i+K_j \leq K_{\rm atm}$, one has $g(m_1=0,K_i,\theta_{13},\d)<4$.
Increasing $m_1$ there is a
suppression due to the fact that $K_i\geq m_1/m_{\star}$ and
$g_{\rm max}(m_1,\theta_{13},\d)$ decreases monotonically. Therefore, for
any $m_1$, there is a lower bound on $M_1$ given by
\be\label{lbM1}
M_1 \geq M_1^{\rm min}(m_1,\theta_{13},\d)\equiv
{3\,\overline{M}_1\over g_{\rm max}(m_1,\theta_{13},\d)}
\,{\d_{j1}\over |\D|}\, .
\ee
The $C\!P$ asymmetries, and consequently the
final asymmetry, are maximally enhanced in the extreme case
of resonant leptogenesis \cite{pilaftsis,pilaund2} when
the heavy neutrino mass degeneracy is comparable to the decay
widths. This implies approximately to have
$\d_{ji}^{\rm res}\simeq d\,\bar{\ve}(M_i)/3$ with $d=1\div 10$,
that would correspond to have $\ve_1=1/d$ in the unflavored case
with maximal phase. This can be taken as a conservative
limit that implies, maximizing over $\d$, a lower bound
\be\label{lbsint}
\sin \theta_{13} \geq
\sin\theta_{13}^{\rm min} = {d\,\eta_B^{\rm CMB}\,N_{\gamma}^{\rm rec}
\over a_{\rm sph}\,
{\rm max}_{\d}[g_{\rm max}(m_1,\theta_{13}^{\rm min},\d)\,\sin\d)]} \, .
\ee
Notice that, within the validity of perturbation theory,
one cannot specify which is the exact value
of $d$, that means the value of $\d_{ji}$
above which the expression for the $C\!P$
asymmetries given in the Eq. (\ref{ve1a}) are valid \cite{abp}
and therefore there is an uncertainty in the calculation of the
maximum enhancement of the asymmetries in the resonant regime.

Let us now specialize the expressions for the four
special cases we have already analyzed in the HL.

\subsection{$M_3\gg 10^{14}\,{\rm GeV}$}

Remember that in this case one has $(h^{\dagger}h)_{3j}=0$ implying
$\ve_{3\a}=0$, a consequence of the fact that the heaviest
RH neutrino decouples. Moreover $m_1\ll m_{\rm sol}$,
such that terms $\propto m_1$ can be neglected, $m_3\simeq m_{\rm atm}$
and $m_2\simeq m_{\rm sol}$ for normal hierarchy or
$m_2\simeq m_{\rm atm}\sqrt{1-m_{\rm sol}^2/m_{\rm atm}^2}$
for inverted hierarchy. Therefore,
there is actually no dependence on $m_1$ in
$g(m_1,\O_{ji},\theta_{13},\d)$ that we can indicate
simply with $g(\O_{ji},\theta_{13},\d)$ and that is given by the
expression (\ref{g}) with $(i,j)=(1,2)$ and $(p,q)=(2,3)$ or
explicitly
\bea \nonumber
g(\O_{21},\theta_{13},\d) & \simeq &
{2\,(K_1+K_2)\,K_{\rm atm}\over K_1\,K_2}\,
\,\left(1-{m_2\over m_{\rm atm}}\right)\,
\sqrt{m_2\over m_{\rm atm}}\,
\O_{21}\,\sqrt{1-\O^2_{21}}\,   \\ \label{gM3}
& & \times
 \sum_{\a}\,\k(K_{1\a}+K_{2\a})\,
{{\rm Im}[U^{\star}_{\a2}\,U_{\a 3}]\over \D }.
\eea
\begin{figure}
\centerline{\psfig{file=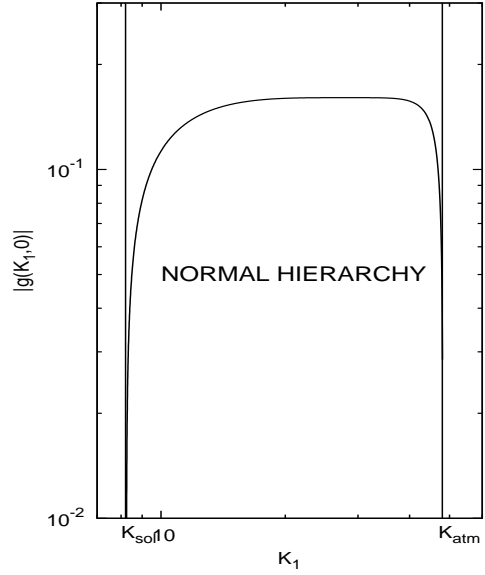,height=80mm,width=70mm,angle=0}}
\caption{Case $M_3\gg 10^{14}\,{\rm GeV}$ for normal hierarchy in the DL.
Plot of the function $|g(K_1,\theta_{13},\d)|$
in the limit $\D\rightarrow 0$.
The maximum gives the lower bound on $M_1$ (see Eq. (\ref{lbM1Minf}))
and on $\sin\theta_{13}$ (see Eq. (\ref{lbsintMinf})).}
\end{figure}
In the case of normal hierarchy $|g(\O_{21},\theta_{13},\d)|$
slightly decreases when $\D$ increases
and so the maximum is found for $\D=0$ and in this case
the dependence on $\theta_{13}$ and on $\d$ disappears.
Replacing $\O_{21}$ with $K_1$, in Fig. 4 we have plotted
$|g(K_1,\D=0)|$ for central values of $m_{\rm sol}$
and $m_{\rm atm}$. Including the errors,
one finds $g_{\rm max}\simeq 0.160 \pm 0.005$.

The ($3\s$) lower bounds on $M_1$ for normal hierarchy,
from the general expression (\ref{lbM1}), is then given by
\be\label{lbM1Minf}
M_1 \geq 0.9 \times \,{10^{10}}\,{\rm GeV}\,{\delta_{21}\over |\D|} \,.
\ee
In the case of inverted hierarchy the situation is somehow opposite,
since for $\theta_{13}=0$ the electron flavor contribution vanishes
in the Eq. (\ref{gM3}) and there is an exact cancellation between the
$\tau$ and $\mu$ contributions. Consequently, the asymmetry
increases for increasing values of $\theta_{13}$ and thus the maximum
is found for $\sin\theta_{13}=0.2$ while $\d\simeq\pi/4$. In this case
one has that
${\rm max}_{\theta_{13},\d}[g_{\rm max}(m_1=0,\theta_{13},\d)\,\D]
\simeq (9\pm 2)\times 10^{-8}$,
that plugged in the Eq. (\ref{lbM1}) gives at $3\s$
\be
M_1\geq 6\times 10^{15}\,{\rm GeV}\,{\delta_{21}}\,.
\ee
It should be remembered that these conditions have been
obtained in the three-flavor regime and in the DL and
therefore are valid for $M_1\lesssim 10^9\,{\rm GeV}$.
This implies $\delta_{21}\lesssim 10^{-1}\,|\D|$ for normal hierarchy
and $\delta_{21}\lesssim 10^{-7}$ for inverted hierarchy.

Analogously the general expression (\ref{lbsint}) gives,
for normal and inverted hierarchy respectively,
the following ($3\s$) lower bounds on $\sin\theta_{13}$:
\be\label{lbsintMinf}
\sin\theta_{13} \gtrsim  3.3\times 10^{-7}\,d
\hspace{5mm}
{\rm and}
\hspace{5mm}
\sin\theta_{13}\gtrsim 0.06\,d .
\ee

\subsection{$\O=R_{13}$}

In this particular case,
the next-to-lightest RH neutrino is decoupled from the other
two and this implies that $\ve_{2\a}=0$ for any $\a$ and that
the $\ve_{1\a}$'s do not depend on $M_2$, in particular they
do not get enhanced if $\d_{21}\rightarrow 0$. Therefore, one
has necessarily to consider $\delta_{31}\lesssim 0.01$,
implying a full DL with all three degenerate RH neutrino masses.
The function $g(m_1,\O_{ji},\theta_{13},\d)$ is now obtained from
the general expression (\ref{g}) for $j=q=3$ and $i=p=1$,
or explicitly
\bea \nonumber
g(m_1,\O_{31},\theta_{13},\d) & \equiv &
{2\,K_{\rm atm}\,(K_1+K_3)\over K_1\,K_3}\,
\,{(m_3-m_1)\,\sqrt{m_3\,m_1}\over m_{\rm atm}^2}\,
\,
\O_{31}\,\sqrt{1-\O_{31}^2} \\ &  & \times   \label{gR13}
\,\sum_{\alpha}\,\k(K_{1\a}+K_{2\a}+K_{3\a})\,
{{\rm Im}[U^{\star}_{\a 1}\,U_{\a 3}]\over\D}\, .
\eea
It is interesting to notice that in this case an
$e$-dominance is realized. Moreover,
one has that the dependence of
$|g(m_1,\O_{31},\theta_{13},\d)|$ on $\theta_{13}$ and $\d$
is slight and the maximum is again for $\D=0$ and for  $m_1=0$
and one finds $g_{\rm max}(0)=0.24\pm 0.01$ for normal hierarchy
and $g_{\rm max}(0)=(3.1\pm 0.2)\times 10^{-3}$ for inverted hierarchy,
so that the general expression (\ref{lbM1}) for the
lower bound on $M_1$ gives, at $3\s$ for normal and inverted hierarchy,
\be
M_1 \gtrsim 5.5\times 10^{9}\,{\rm GeV}\,{\delta_{31}\over |\D|}
\hspace{5mm}
{\rm and}
\hspace{5mm}
M_1 \gtrsim 5 \times 10^{11}\,{\rm GeV}\,{\delta_{31}\over |\D|}
\, ,
\ee
while the general expression (\ref{lbsint}) in resonant leptogenesis gives
\be
\sin\theta_{13}\gtrsim 2.3\times 10^{-7}\,d
\hspace{5mm}
{\rm and}
\hspace{5mm}
\sin\theta_{13}\gtrsim 1.5\times 10^{-5}\,d \, .
\ee
Increasing $m_1$, the value of $g_{\rm max}(m_1)$ decreases and
the lower bound on $\sin\theta_{13}$ in resonant leptogenesis
becomes more and more
restrictive. This dependence is shown in Fig. 5 both for normal
(left panel) and inverted (right panel) hierarchy and for
$d=1$ (solid line) and $d=10$ (short-dashed line).
Very interestingly, imposing
the experimental ($3\s$) upper bound $\sin\theta_{13}\lesssim 0.20$,
one obtain the upper bound $m_1 \lesssim \, (0.2-0.4)\,{\rm eV}$,
depending on the value of $d$.
\begin{figure}
\psfig{file=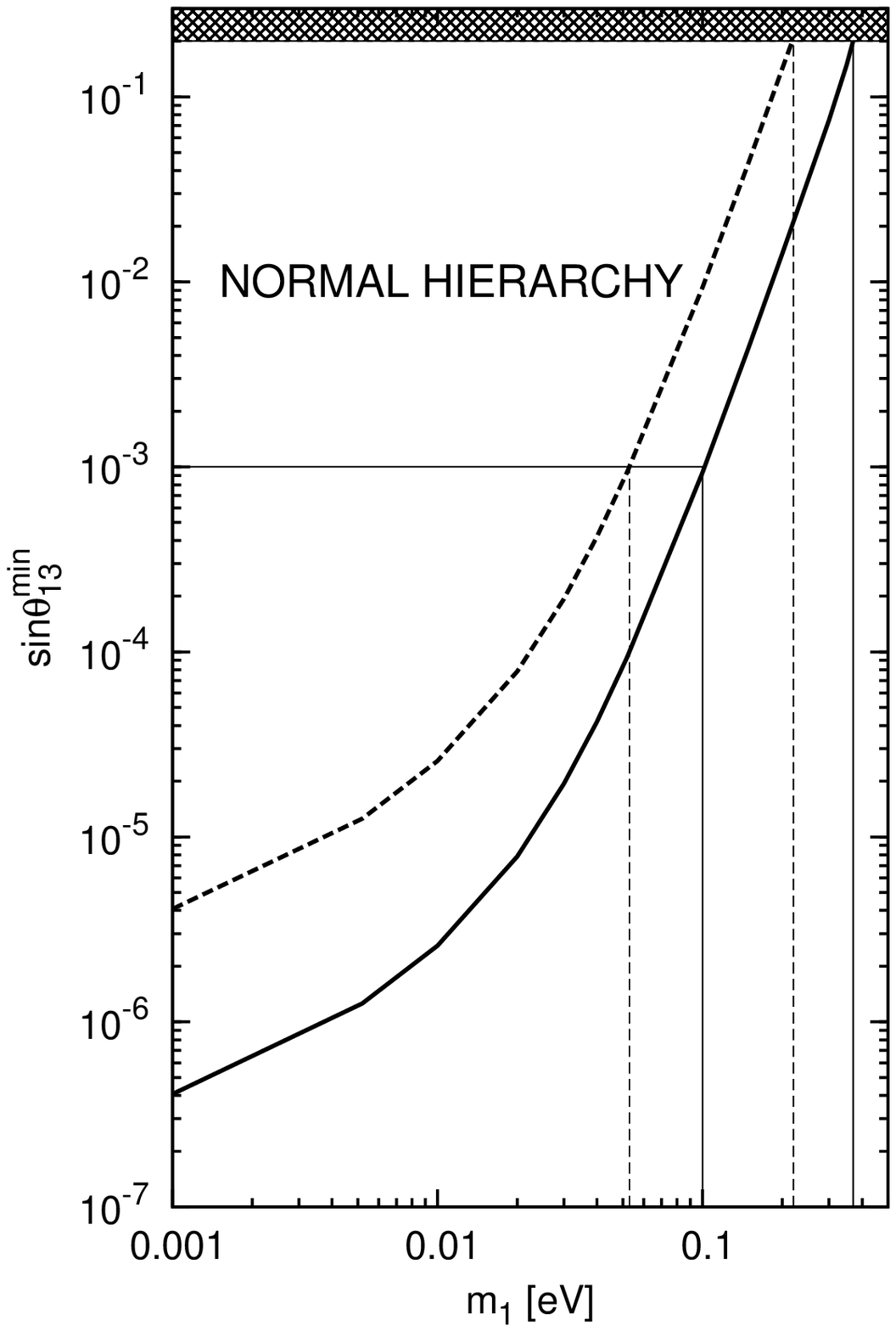,height=80mm,width=70mm,angle=0}
\hspace{5mm}
\psfig{file=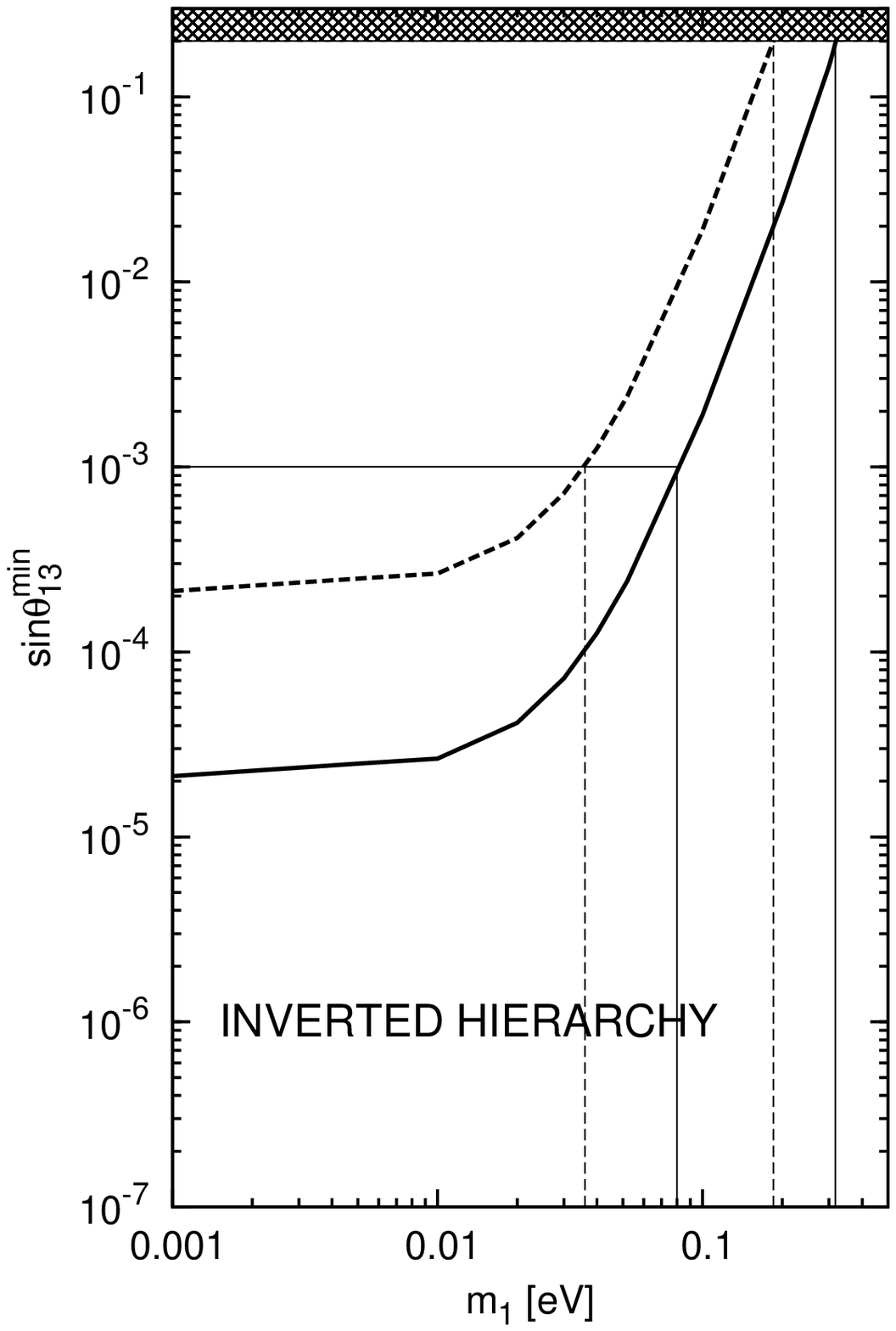,height=80mm,width=70mm,angle=0}
\caption{Case $\O=R_{13}$ in the full DL. Lower bound on
$\sin\theta_{13}$ versus $m_1$ obtained in resonant leptogenesis
for $d=1$ (solid line) and $d=10$ (short-dashed line).
Values $\sin\theta_{13} > 0.20$ are excluded at $3\,\s$ by current
experimental data.}
\end{figure}
This upper bound will become more stringent if the experimental
upper bound on $\sin\theta_{13}$ will improve, as expected
in future experiments in the case of no discovery.
The most stringent experimental
upper bound that can be hopefully reached in future with
neutrino factories is approximately
$\sin\theta_{13}< 10^{-3}$ \cite{lindner}. This asymptotical
upper bound is also shown in Fig. 5 and would imply
an upper bound $m_1 \lesssim \, (0.05-0.1)\,{\rm eV}$
for normal hierarchy and $m_1 \lesssim \, (0.03-0.08)\,{\rm eV}$
for inverted hierarchy. Therefore,
an interesting interplay between two measurable quantities
is realized and this makes $\d$-leptogenesis falsifiable independently
of the RH neutrino mass spectrum.

In the more conservative case of normal hierarchy,
see left panel of Fig.~5, a good approximation
is given by the fit
\be
m_1\lesssim 0.6\,
\left({\sin\theta_{13}- 2.3\times 10^{-7}} \right)^{0.25} \,{\rm eV} \, .
\ee
It is interesting that this upper bound
holds in the extreme case of resonant leptogenesis and
therefore holds for any RH neutrino spectrum. However,
we have to verify whether it holds also for a different
choice of $\O$.

\subsection{$\O=R_{12}$}

The situation for $\O=R_{12}$ is quite different
compared to the previous cases. Now one has $i=p=1$
and $j=q=2$ and it is possible to have both a partial
DL with $10^{14}\,{\rm GeV}\gtrsim M_3 \gg M_2\simeq M_1$
and a full DL. In the first case, the general expression
Eq. (\ref{g})  becomes
\bea \nonumber
g(m_1,\O_{21},\theta_{13},\d) & \equiv &
{2\,K_{\rm atm}\,(K_1+K_2)\over K_1\,K_2}\,
\,{(m_2-m_1)\,\sqrt{m_2\,m_1}\over m_{\rm atm}^2}\,
\,
\O_{21}\,\sqrt{1-\O_{21}^2} \\ &  & \times   \label{gR21}
\,\sum_{\alpha}\,\k(K_{1\a}+K_{2\a})\,
{{\rm Im}[U^{\star}_{\a 1}\,U_{\a 2}]\over\D}\, .
\eea
\begin{figure}
\psfig{file=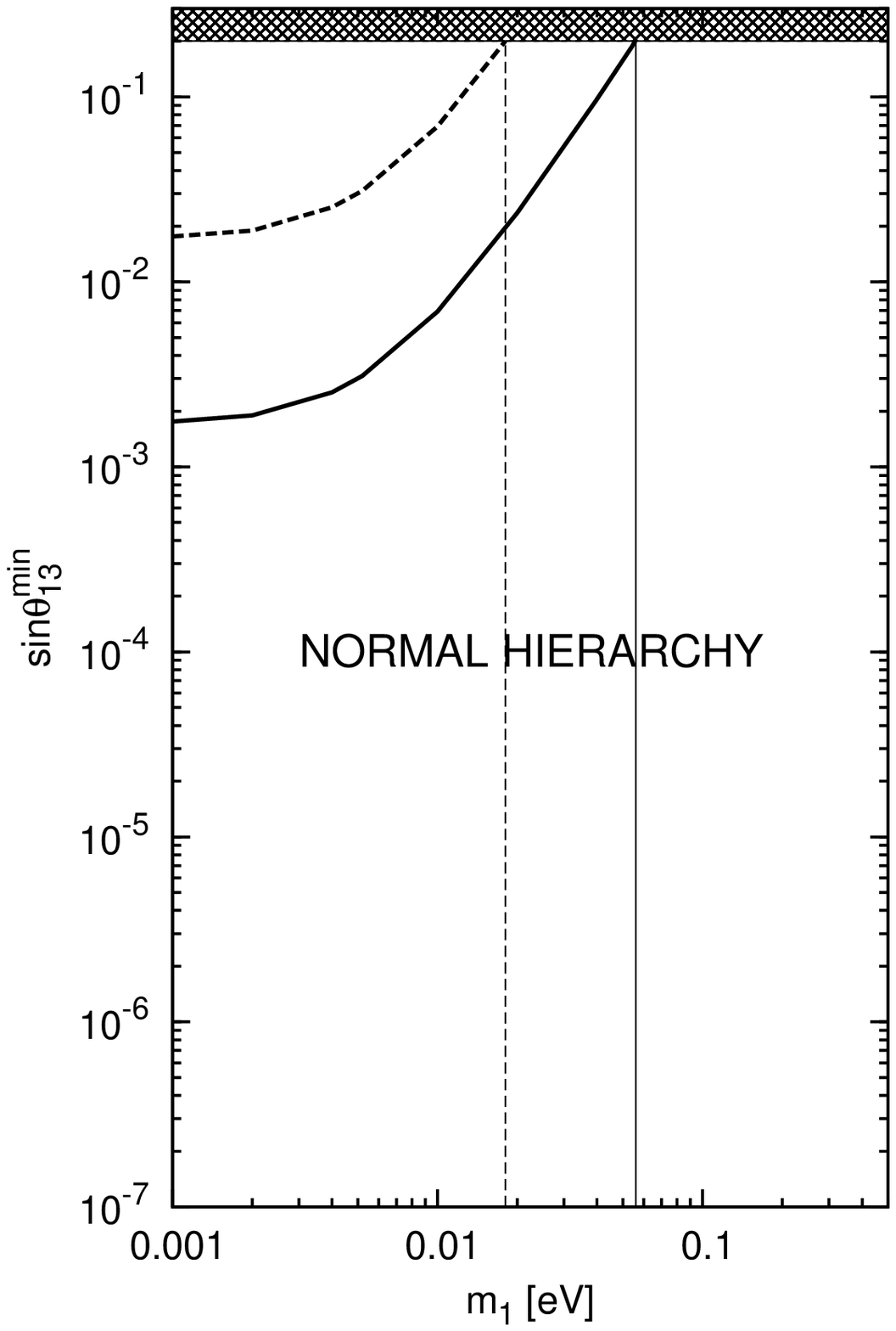,height=80mm,width=70mm,angle=0}
\hspace{5mm}
\psfig{file=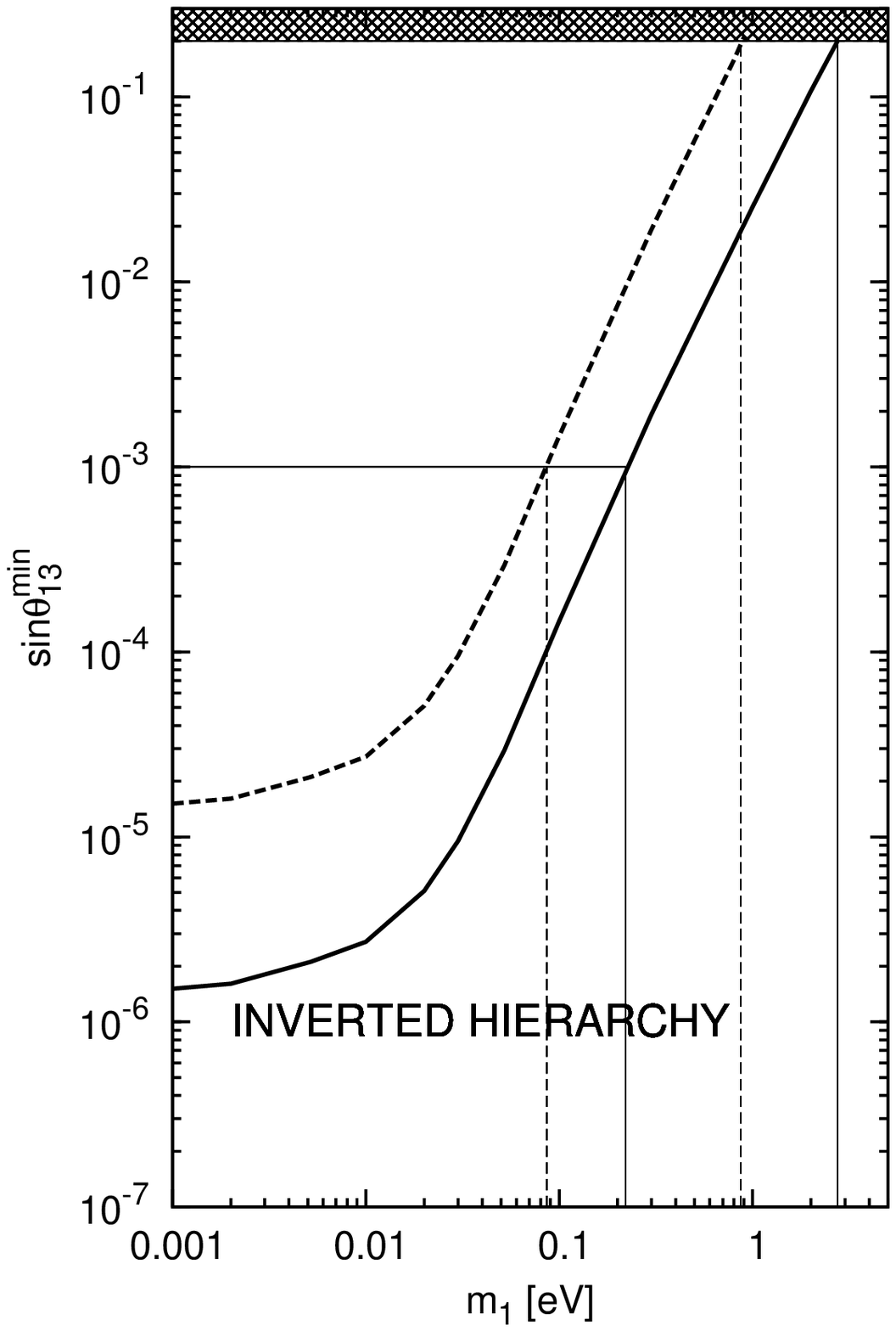,height=80mm,width=70mm,angle=0}
\caption{Case $\O=R_{12}$ in the partial DL. Lower bound on
$\sin\theta_{13}$ versus $m_1$ obtained in resonant leptogenesis.
Same conventions as in the previous figure.}
\end{figure}
This time the contribution from the electron flavor vanishes.
Furthermore, for normal hierarchy, there is an almost perfect cancellation
between the $\m$ and the $\t$ contribution. In the left panel
of Fig. 6 we show the lower bound on $\sin\theta_{13}$ versus $m_1$ and
one can see how, compared to the previous case $\O=R_{13}$, this is
much more restrictive. In particular, imposing $\sin\theta_{13}<0.2$,
one obtains now a much more stringent upper bound $m_1\lesssim 0.06\,{\rm eV}$.
On the other hand, for inverted hierarchy,
the cancellation between the $\m$ and the $\tau$ flavor does not occur
and one has a lower bound on $\sin\theta_{13}$, for $m_1\ll 0.01\,{\rm eV}$,
shown in the right panel of Fig. 6, that is very similar to what has been
obtained in the case $\O=R_{13}$. However, now there is no
flavor cancellation for increasing values of $m_1$,
because $K_{1\a}+K_{2\a}$ does not tend to a common value like
$\sum_j\,K_{j\a}$. Therefore, one can see in Fig. 6 that this time
the upper bound on $m_1$ is much looser, both compared to normal
hierarchy and compared to $\O=R_{13}$.

In the full DL, the flavor
cancellation at large $m_1$ occurs and the results are shown
in Fig. 7. One can see how now for normal hierarchy the upper bound
on $m_1$ is even much more restrictive and, for inverted hierarchy, one has
a situation that is similar to the case $\O=R_{13}$.
\begin{figure}
\psfig{file=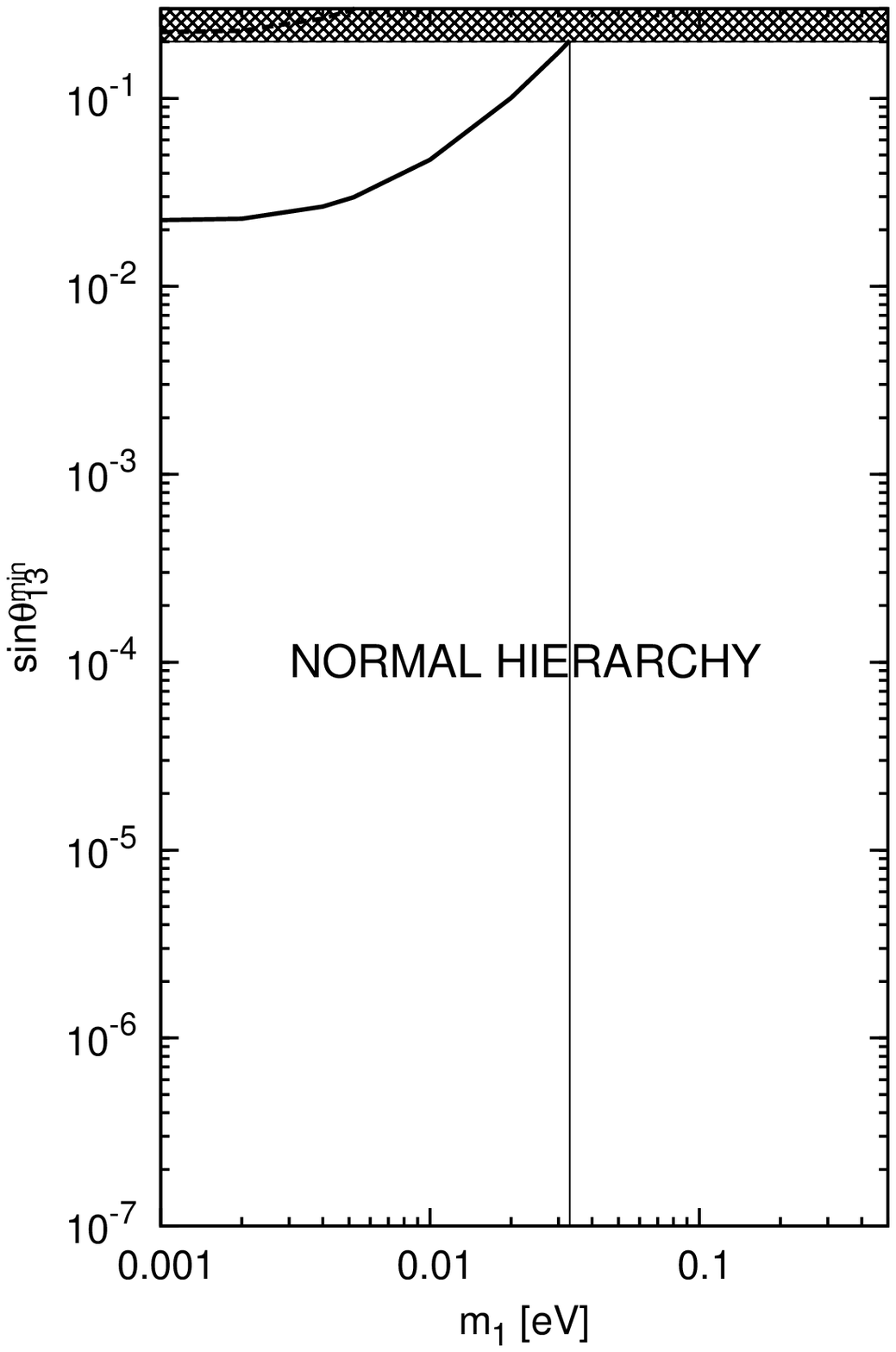,height=80mm,width=70mm,angle=0}
\hspace{5mm}
\psfig{file=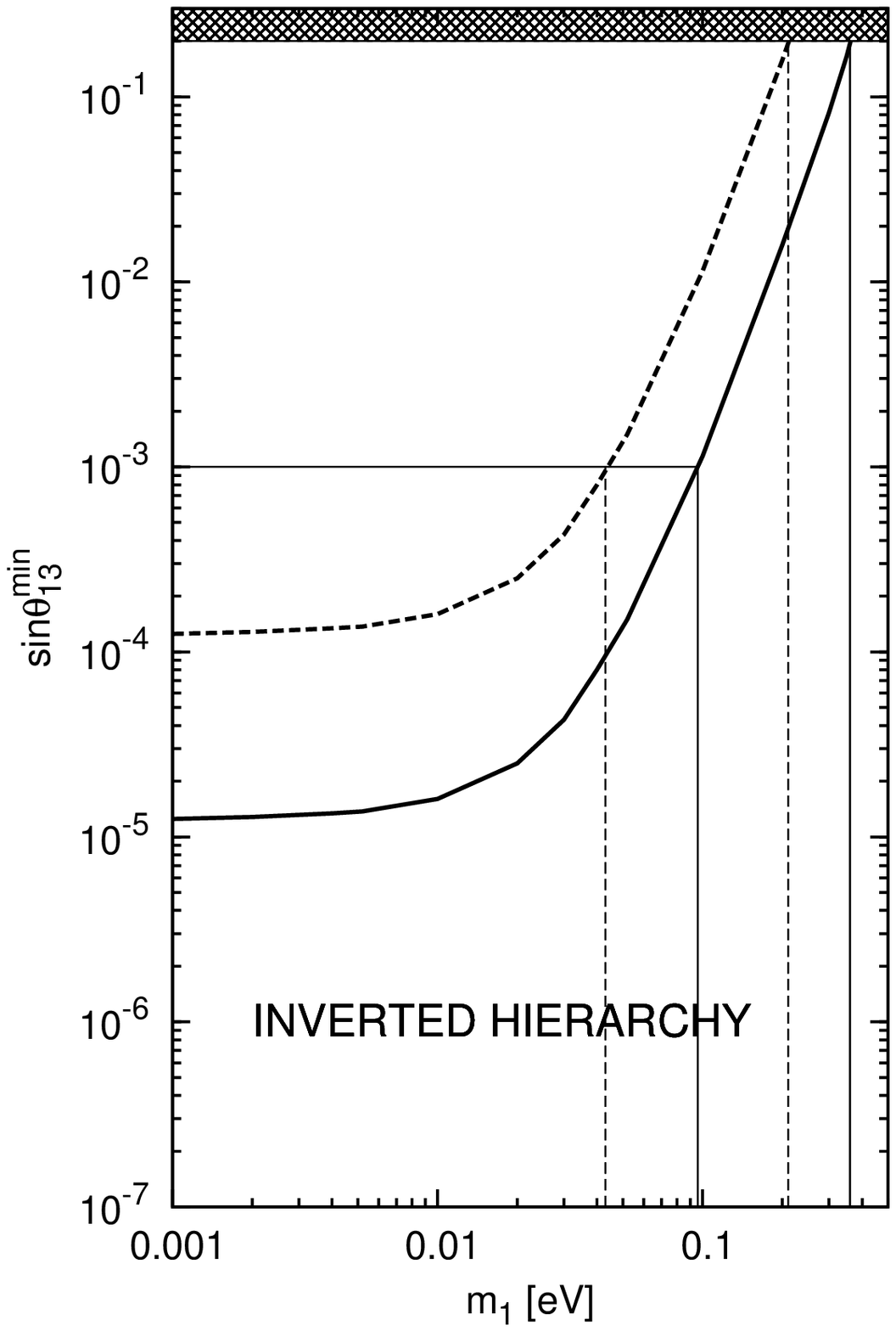,height=80mm,width=70mm,angle=0}
\caption{Case $\O=R_{12}$ in the full DL. Lower bound on
$\sin\theta_{13}$ versus $m_1$ obtained in resonant leptogenesis.
Same conventions as in the previous figures.}
\end{figure}

\subsection{$\O=R_{23}$}

In this case the lightest RH neutrino decouples and
$\ve_{1\a}=0$, independently of $M_1$.
Therefore, there is no contribution to the final asymmetry
from $N_1$ decays. On the other hand $\ve_{2\a}$ and $\ve_{3\a}$
do not vanish and therefore there is a contribution from the
decays of the two heavier RH neutrinos. Still $N_1$ inverse
processes have to be taken into account since they
contribute to the wash-out.
There are two different possibilities.

In a full DL the wash-out from $N_1$ inverse decays just cumulates
with the wash-out from the two heavier. Therefore, this time,
in the expression Eq. (\ref{g}), one has $i=p=2$ and $j=q=3$
and $\k_{\a}^{\rm f}=\k(K_{1\a}+K_{2\a}+K_{3\a})$, explicitly
\bea \nonumber
g(m_1,\O_{32},\sin\theta_{13},\sin\d) & \equiv &
{2\,K_{\rm atm}\,(K_2+K_3)\over K_2\,K_3}\,
\,{(m_3-m_2)\,\sqrt{m_3\,m_2}\over m_{\rm atm}^2}\,
\,
\O_{32}\,\sqrt{1-\O_{32}^2} \\ &  & \times   \label{g32a}
\,\sum_{\alpha}\,\k(K_{1\a}+K_{2\a}+K_{3\a})\,
{{\rm Im}[U^{\star}_{\a 2}\,U_{\a 3}]\over\D}\, .
\eea
In Fig. 8 we show the dependence of the $\sin\theta_{13}$ lower bound
on $m_1$. This time there is a bigger suppression than in the
case $\O=R_{13}$, both for normal and for inverted hierarchy.
\begin{figure}
\psfig{file=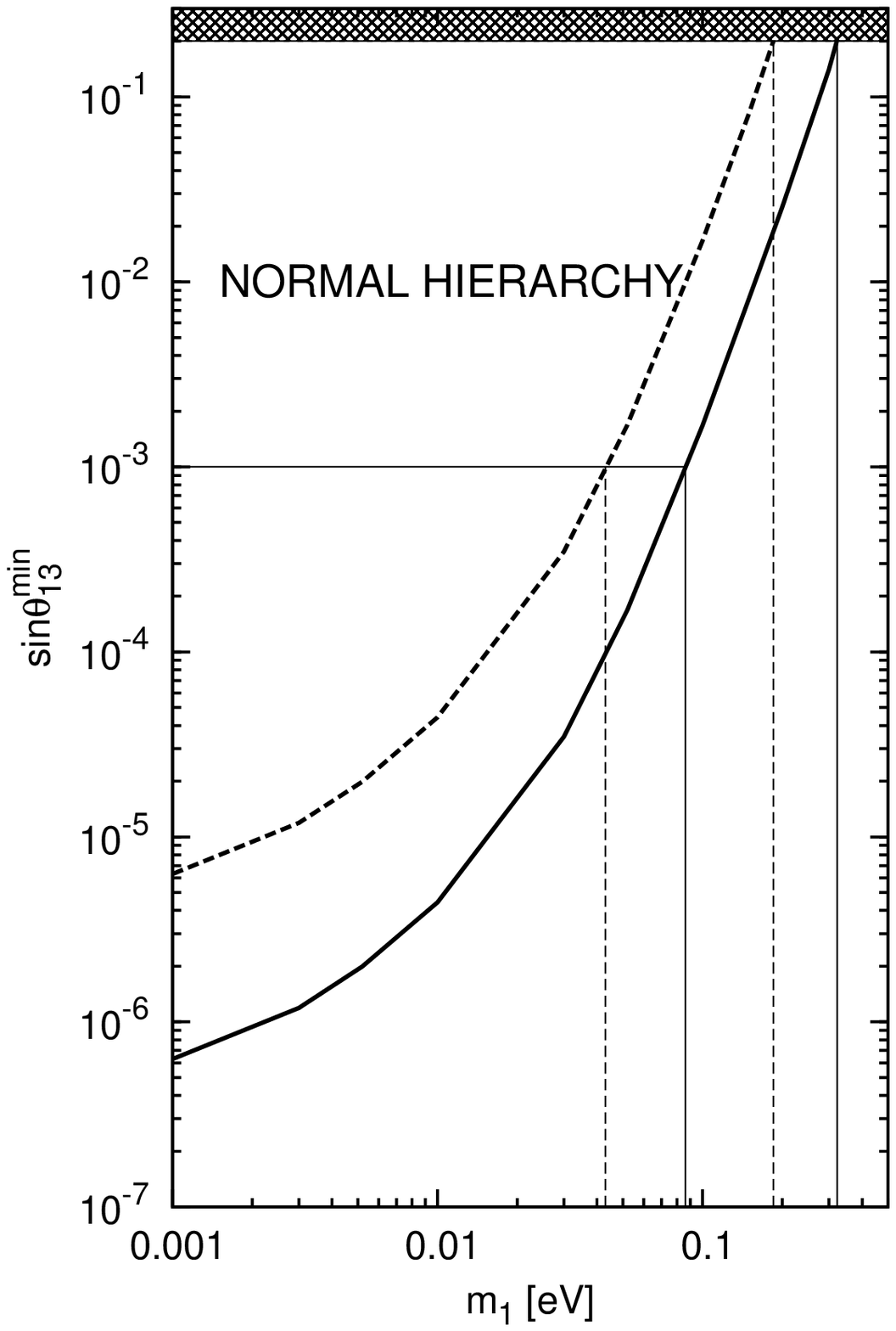,height=80mm,width=70mm,angle=0}
\hspace{5mm}
\psfig{file=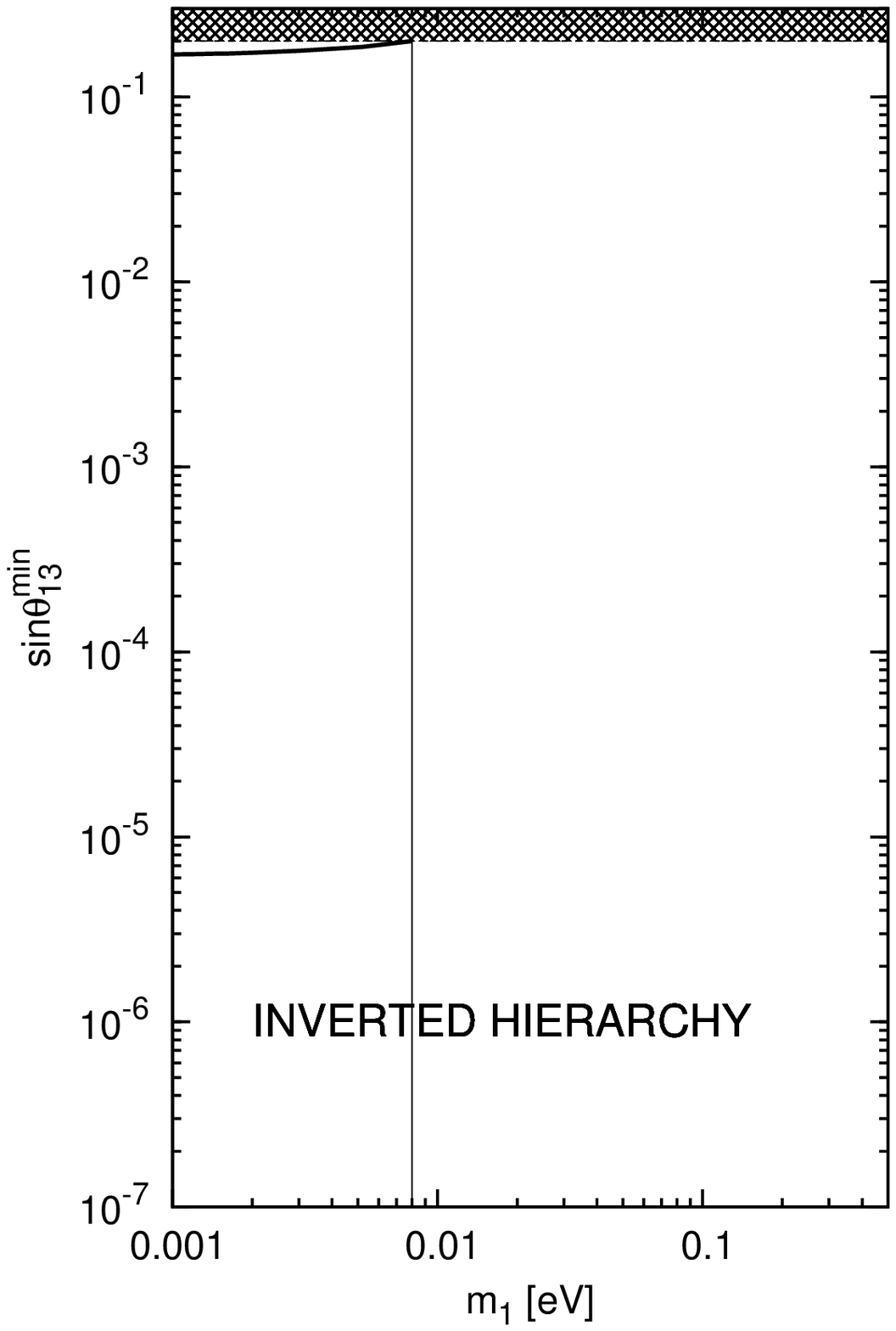,height=80mm,width=70mm,angle=0}
\caption{Case $\O=R_{23}$ in the full DL.
Lower bound on $\sin\theta_{13}$ versus
$m_1$ obtained in resonant leptogenesis.
Same conventions as in the previous figures.}
\end{figure}

In the case $M_1\ll M_2\simeq M_3$ one has
\bea \nonumber
g(m_1,\O_{32},\sin\theta_{13},\sin\d) & \equiv &
{2\,K_{\rm atm}\,(K_2+K_3)\over K_2\,K_3}\,
\,{(m_3-m_2)\,\sqrt{m_3\,m_2}\over m_{\rm atm}^2}\,
\,
\O_{32}\,\sqrt{1-\O_{32}^2} \\ &  & \times   \label{g32b}
\,\sum_{\alpha}\,\k(K_{2\a}+K_{3\a})
\,e^{-{3\,\pi\over 8}\,K_{1\a}}
{{\rm Im}[U^{\star}_{\a 2}\,U_{\a 3}]\over\D}\, .
\eea
The dependence of the lower bound on $\sin\theta_{13}$ on $m_1$
is shown in Fig. 9 for normal hierarchy. In this case
the upper bound on $m_1$ is now slightly less stringent than
in the previous cases.
\begin{figure}
\centerline{
\psfig{file=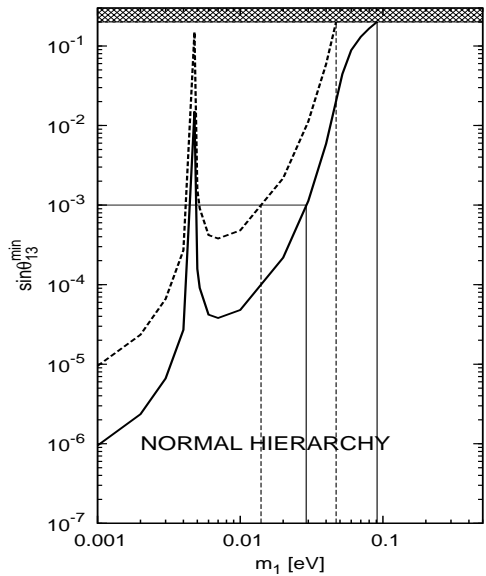,height=80mm,width=70mm,angle=0}}
\caption{Case $\O=R_{23}$ in the partial DL.
Lower bound on $\sin\theta_{13}$ versus $m_1$ obtained in resonant
leptogenesis. Same conventions as in the previous figures.}
\end{figure}
For inverted hierarchy the final asymmetry production is so
suppressed that there is no allowed region.

We can conclude this section noticing that these results
show that $\d$-leptogenesis can be falsified. In the case
of normal hierarchy, the current upper bound
$\sin\theta_{13}\lesssim 0.2$ implies
$m_1\lesssim 0.1\,{\rm eV}$, while, in future, a
potential upper bound
$\sin\theta_{13}\lesssim 10^{-3}$ would imply
$m_1\lesssim {\cal O}(0.01\,{\rm eV})$, with
a more precise determination depending on the possibility
of improving the current estimation of the parameter $d$ in
resonant leptogenesis.

\section{Lights and shadows of $\d$-leptogenesis}

The most attractive feature of $\d$-leptogenesis is that
a non-vanishing Dirac phase, the only see-saw phase
that we can realistically hope to discover in future,
acts as the only source of $C\!P$ violation responsible
for the matter-antimatter asymmetry of the Universe.
We think that this feature, despite of the objections
that we are going to discuss,
provides a strong motivation for $\d$-leptogenesis.

As we have seen, successful $\d$-leptogenesis implies
stringent conditions on the RH neutrino masses, something quite
interesting since they escape conventional experimental information.
In particular we have seen that, except for a marginal
allowed region in the weak wash-out regime,
the HL is non-viable. We also observed
that a definite conclusion on the existence of such a marginal allowed
region, requires a full quantum kinetic treatment but in any case
corrections are expected to shrink this already quite
restricted allowed region.

Therefore, $\d$-leptogenesis motivates models
with degenerate RH neutrino masses, with
the most extreme limit represented by resonant leptogenesis.
Even in this extreme limit however,  imposing successful $\d$-leptogenesis,
interesting conditions follow on quantities accessible in low-energy
neutrino experiment: $\sin\theta_{13}$, the absolute neutrino mass
scale, normal or inverted scheme, the Dirac phase itself.
Therefore, an interesting aspect of $\d$-leptogenesis is that it is
falsifiable independently of the heavy neutrino mass spectrum.

There are some objections to $\d$-leptogenesis.
There is no clear theoretical motivation for
$\d$-leptogenesis, more generally
to choose a real orthogonal $\O$ matrix. Apparently,
sequential dominated models \cite{king} could represent
an interesting theoretical framework. Indeed in \cite{geometry}
it was shown that these models correspond
to have an $\O$ matrix that slightly deviates from the
unit matrix or from all the other five that can be
obtained from the unit matrix exchanging rows or columns.
However, it has been noticed \cite{window,geometry} that
in the limit ${\rm Im}[\O]\rightarrow 0$
total $C\!P$ asymmetries $\ve_i$ do not necessarily vanish. Therefore,
in this limit and taking vanishing Majorana phases,
one does not necessarily obtain $\d$-leptogenesis. Writing
$\O^2_{ij}=|\O^2_{ij}|\,{\rm exp}[i\,\varphi_{ij}]$, the correct condition
to enforce $\ve_{i}\rightarrow 0$ is to take the limit
$\varphi_{ij}\rightarrow 0$. This is a more demanding limit
than ${\rm Im}[\O]\rightarrow 0$ and it is currently
not motivated by generic sequential dominated models.
This limit is not motivated either by radiative leptogenesis
\cite{radiative} within the context of
the minimal flavor violation principle \cite{MFV},
as recently considered in \cite{burasbranco,selma}.
Therefore, there is no theoretical justification
for $\d$-leptogenesis at the moment.

Another possible objection to $\d$-leptogenesis is that
it cannot be distinguished from the general scenario,
where all phases are present, even if a non-vanishing
Dirac phase is discovered. Indeed a Dirac phase would
give in this case a subdominant contribution.
This objection is however related also to the first one.
Indeed, since a theoretical model motivating $\d$-leptogenesis
is required anyway, one can hope to find some specific
prediction that makes the model testable and $\d$-leptogenesis
together with it. Dirac phase leptogenesis would then become
distinguishable from the general scenario, though in an indirect way.

This last objection can be also considered
within a more particular case where $\O$ is still real but
Majorana phases are present together with the
Dirac phase. It has been noticed that the contribution to the final asymmetry
from Majorana phases is in general dominant compared to that
one coming from the Dirac phase \cite{flavorlep}.
In the right panel of Fig. 1
we have compared the result on the $M_1$
lower bound for $\O=R_{13}$ obtained in $\d$-leptogenesis with the
result when ${\rm Im}[\O_{ij}]=\d=0$ but $\Phi_1=-\pi/2$ (dotted lines).
One can see that in the second case the lower bound is $\sim 2\div 3$
times more relaxed.
This result can be easily understood analytically \cite{pascoli}
and actually it can be also observed that there can be
exact cancellations between the contribution to the final asymmetry
from the Majorana phases and from the Dirac phase.

The presence of cancellations
can be somehow regarded as a limit to $\d$-leptogenesis main
motivation, since even though a Dirac phase will be discovered,
it is not guaranteed that the observed asymmetry can be explained.
This objection is however quite weak since it would be quite strange if Nature
disposed a  sufficient source of $C\!P$ violation but
set up a second source that exactly cancels with the first one while
the observed asymmetry is, in the end, explained still by a third one,
for example the phases in $\O$. On the other hand, we can say that
it would be certainly positive for $\d$-leptogenesis if in future
experimental upper bounds on the Majorana phases are placed,
for example from $\b\b0\n$ decay, thus constraining the
contribution to the final asymmetry from Majorana phases \cite{pascoli}.
This can be also regarded as a further prediction coming from $\d$-leptogenesis.

In conclusion, we have studied in detail a specific scenario of
leptogenesis that is interesting especially in
view of the many next planned experiments aiming at a discovery of
$C\!P$ violation in neutrino mixing. Despite some important
remarks and objections, we think that $\d$-leptogenesis realizes
a very interesting link between a long-standing cosmological puzzle
and $C\!P$ violation in neutrino oscillations,
one of the most relevant experimental topics in high-energy
physics during next years.

\vspace{2mm}
\noindent
\textbf{Acknowledgments}\\
It is a pleasure to thank  S.~Petcov for discussions during the Neutrino
Oscillation Workshop 2006, held in Conca  della Specchiulla (Italy).
We also wish to thank G.~Raffelt for interesting discussions and
comments.
This work was supported, in part, under the Marie Curie project
``Leptogenesis, Seesaw and GUTs,'' contract No.~MEIF-CT-2006-022950.

\end{document}